\documentclass[aps,prd,twocolumn,superscriptaddress,
floatfix,nofootinbib]{revtex4-1}

\pdfoutput=1

\usepackage[dotinlabels]{titletoc}
\usepackage{titlesec}

\usepackage{xcolor}
\usepackage{mathrsfs}
\usepackage{amsmath}
\usepackage{amssymb}
\usepackage{bm}
\usepackage{amsfonts}
\usepackage{subfigure}
\usepackage{feynmp}
\usepackage{extarrows}
\usepackage{slashed}
\usepackage{graphicx}
\usepackage{dcolumn}
\usepackage{verbatim}
\usepackage{here}
\usepackage{multirow} 
\allowdisplaybreaks[4]
\usepackage{indentfirst}
\usepackage{isodateo}
\usepackage[low-sup]{subdepth}

\definecolor{gesfpurple}{rgb}{0.47,0.19,0.42}

\definecolor{gesflanse}{rgb}{0.00,0.50,0.50}

\definecolor{gesfblue}{rgb}{0.08,0.42,0.76}

\definecolor{gesfred}{rgb}{1,0,0}

\definecolor{gesfwhite}{rgb}{1,1,1}

\definecolor{gesfblack}{rgb}{0,0,0}

\makeatletter
\newsavebox\myboxA
\newsavebox\myboxB
\newlength\mylenA
\newcommand*\xoverline[2][0.7]{%
	\sbox{\myboxA}{$\m@th#2$}%
	\setbox\myboxB\null
	\ht\myboxB=\ht\myboxA%
	\dp\myboxB=\dp\myboxA%
	\wd\myboxB=#1\wd\myboxA
	\sbox\myboxB{$\m@th\overline{\copy\myboxB}$}
	\setlength\mylenA{\the\wd\myboxA}
	\addtolength\mylenA{-\the\wd\myboxB}%
	\ifdim\wd\myboxB<\wd\myboxA%
	\rlap{\hskip 0.8\mylenA\usebox\myboxB}{\usebox\myboxA}%
	\else
	\hskip -0.5\mylenA\rlap{\usebox\myboxA}{\hskip 0.5\mylenA\usebox\myboxB}%
	\fi}

\usepackage{isodateo}
\usepackage[CJKbookmarks=true, bookmarksnumbered=true,bookmarksopen=true,]{hyperref}
\hypersetup{colorlinks,%
              linkcolor=blue,
              citecolor=blue,
              urlcolor=blue}
\graphicspath{{figs/}}
\usepackage{cleveref}
\crefname{equation}{Eq.}{Eqs.\!} 
\crefrangelabelformat{equation}{(#3#1#4--#5#2#6)}

\definecolor{Green}{RGB}{199,238,206}

\newcommand{\eqrefe}{Eq.\eqref}
\newcommand{\beq}{\begin{equation}}
\newcommand{\eeq}{\end{equation}}
\newcommand{\ba}{\begin{array}}
\newcommand{\ea}{\end{array}}
\newcommand{\beqa}{\begin{eqnarray}}
\newcommand{\eeqa}{\end{eqnarray}}
\newcommand{\beqs}{\begin{subequations}}
\newcommand{\eeqs}{\end{subequations}}

\newcommand{\la}{\langle}
\newcommand{\ra}{\rangle}
\newcommand{\fr}[2]{\mbox{$\frac{\,{#1}\,}{#2}$}}
\renewcommand{\rm}{\mathrm}
\def\LB{\left[}

\def\leqq{\leqslant}
\def\geqq{\geqslant}
\def\({\left(}
\def\){\right)}
\def\[{\left[\,}
\def\]{\,\right]}
\def\LB{\left\{}

\def\nn{\nonumber}
\def\pd{\partial}

\def\pp{\prime}
\def\to{\rightarrow}
\def\ito{\!\rightarrow\!}

\def\TT{\mathcal{T}}

\def\dT{\delta\mathcal{T}}
\def\M{\mathcal{M}}
\def\MT{\widetilde{\mathcal{M}}}
\def\dM{\delta\mathcal{M}}
\def\dMT{\delta\widetilde{\mathcal{M}}}

\def\zT{\tilde{z}}

\def\NN{\mathcal{N}}
\def\NNt{\widetilde{\mathcal{N}}}
\def\dNN{\delta\mathcal{N}}
\def\dNNt{\delta\widetilde{\mathcal{N}}}

\def\DEn{\widetilde{\Delta}_n^{}}

\def\An{\mathcal{A}_n^\mu}

\def\phin{\phi_n^{}}

\def\A{\mathcal{A}}

\def\CC{\mathcal{C}}
\def\D{\mathcal{D}}

\def\La{\mathcal{L}}
\def\mO{\mathcal{O}}

\def\T{\mathcal{T}}
\def\tT{\widetilde{\mathcal{T}}}

\def\hh{\hat{h}}
\def\hg{\hat{g}}
\def\phih{\hat{\phi}}
\def\ZZ{\mathbb{Z}_2^{}}
\def\ii{\text{i}}
\def\thn{\tilde{h}_n}

\def\al{\alpha}

\def\be{\beta}

\def\ka{\kappa}
\def\ab{\alpha\beta}

\def\mn{\mu\nu}

\def\dnm{\delta_{nm}^{}}
\def\da{\delta}
\def\td{\text{d}}
\def\heta{\hat{\eta}}
\def\ep{\epsilon}
\def\vep{\varepsilon}
\def\lam{\lambda}

\def\hka{\hat{\kappa}}

\def\ba{\bar{a}}

\def\bs{\bar{s}}

\def\vnt{\tilde{v}_n^{}}
\def\vt{\tilde{v}}
\def\tX{\widetilde{X}}
\def\ct{c_\theta}
\def\st{s_\theta}

\def\ctt{c_{2\theta}}
\def\cttt{c_{3\theta}}
\def\ctf{c_{4\theta}}
\def\ctfif{c_{5\theta}}
\def\cts{c_{6\theta}}

\def\diag{\text{diag}}
\def\Rxi{R_{\xi}^{}}
\def\sz{s_0^{}}

\def\tz{t_0^{}}
\def\uz{u_0^{}}
\def\hLn{h_L^n}

\def\Afn{A_5^n}
\def\Mn{M_n^{}}
\def\Mnn{M_n^2}
\def\EE{\mathcal{E}}
\def\VV{\mathcal{V}}
\def\Sb{\mathbb{S}}
\def\MD{\mathcal{M}_{\Delta}^{}}

\def\pnnnn{\phi_{n} \phi_{n} \!\ito \phi_{n}  \phi_{n} }

\def\hs{\hspace*{0.3mm}}
\def\hsx{\hspace*{0.5mm}}
\def\hsm{\hspace*{-0.3mm}}
\def\hsmx{\hspace*{-0.5mm}}

\def\fP{\mathsf{P}}

\allowdisplaybreaks
\begin{document}

\title{Gravitational Equivalence Theorem and Double-Copy
\\[1mm]
for Kaluza-Klein Graviton Scattering Amplitudes}

\author{{\sc Yan-Feng Hang}}
\email[]{yfhang@sjtu.edu.cn}
\affiliation{%
Tsung-Dao Lee Institute $\&$ School of Physics and Astronomy,\\
Key Laboratory for Particle Astrophysics and Cosmology (MOE),\\
Shanghai Key Laboratory for Particle Physics and Cosmology,\\
Shanghai Jiao Tong University, Shanghai, China}

\author{\,{\sc Hong-Jian He}}
\email[]{hjhe@sjtu.edu.cn}
\affiliation{%
Tsung-Dao Lee Institute $\&$ School of Physics and Astronomy,\\
Key Laboratory for Particle Astrophysics and Cosmology (MOE),\\
Shanghai Key Laboratory for Particle Physics and Cosmology,\\
Shanghai Jiao Tong University, Shanghai, China}
\affiliation{%
Institute of Modern Physics $\&$ Physics Department,
Tsinghua University, Beijing, China;
\\
Center for High Energy Physics, Peking University, 
Beijing, China
}


\begin{abstract} 	
\noindent
We analyze the structure of scattering amplitudes of the
Kaluza-Klein (KK) gravitons and
of the KK gravitational Goldstone bosons
in the compactified 5d General Relativity (GR).
Using a general $\Rxi$ gauge-fixing,
we study the geometric Higgs mechanism 
for the massive spin-2 KK gravitons.\ 
We newly propose and prove a Gravitational Equivalence Theorem
(GRET) to connect the scattering amplitudes of longitudinal KK
gravitons to that of the KK gravitational Goldstone bosons,
which {\it formulates the geometric gravitational Higgs mechanism 
at the scattering $S$-matrix level.} 
We demonstrate that {\it the GRET provides a general energy-cancellation mechanism} guaranteeing the $N$-point
longitudinal KK graviton scattering amplitudes
to have their leading energy dependence cancelled down
by a large power factor of $E^{2N}$\,($N\!\geqq\! 4$) 
up to any loop order.\ 
We propose an extended double-copy approach to construct 
the massive KK graviton (Goldstone) amplitudes from
the KK gauge boson (Goldstone) amplitudes.\ 
With these we establish {\it a new correspondence between the
two types of energy cancellations} in the
four-point longitudinal KK amplitudes at tree level:\
$\,E^4\!\ito E^0\,$ in the KK gauge theory
and $\,E^{10}\!\!\to\! E^2\,$ in the KK GR theory.
\\[2mm]
{Journal-Ref: {\tt Research}, vol.\,2022 (2022), Article\,ID 9860945.}
\\[1mm]
{\url{https://doi.org/10.34133/2022/9860945}}
\end{abstract}

\maketitle	


\section{\hspace*{-2.5mm}Introduction}
\label{sec:1}
\vspace*{-3mm}
%

Kaluza-Klein (KK) compactification\,\cite{KK}
of the extra spatial dimensions
leads to infinite towers of massive KK excitation states
in the low energy 4d effective field theory.\ 
This serves as an essential ingredient
of all extra dimensional models\,\cite{Exd} and the string/M  theories\,\cite{string}.\
The KK compactification realizes
the geometric ``Higgs'' mechanisms
for mass generations of KK gravitons\,\cite{GHiggs}
and of KK gauge bosons\,\cite{5DYM2002}
without invoking any extra Higgs boson
of the conventional Higgs mechanism\,\cite{higgsM}.

\vspace*{1mm}

In this work, we formulate the geometric gravitational
``Higgs'' mechanism for the compactified 5d General Relativity (GR5)
by quantizing the KK GR5 under
a general $R_\xi^{}$ gauge-fixing at both the Lagrangian level
and the $S$-matrix level.\ We prove that the KK graviton propagator
is {\it free from the longstanding problem} of 
van\,Dam-Veltman and Zakharov (vDVZ) discontinuity\,\cite{vDVZ} 
in the conventional Fierz-Pauli massive gravity\,\cite{PF}\cite{Hinterbichler:2012} 
and the KK GR5 theory can {\it consistently 
realize the mass-generation for spin-2 KK gravitons.}\ 
Then, we propose and prove 
a new Gravitational Equivalence Theorem (GRET)
which quantitatively connects each scattering amplitude
of the (helicity-zero) longitudinally-polarized KK gravitons 
to that of the corresponding KK Goldstone bosons.\
The GRET takes a highly nontrivial form and
differs substantially from the KK Gauge Equivalence Theorem (GAET)
of the 5d KK gauge
theories\,\cite{5DYM2002}\cite{5DYM2002-2}\cite{KK-ET-He2004},
because each massive KK graviton $h_n^{\mn}$
has 5 helicity states ($\lambda\!\hsm =\!0,\pm 1,\pm 2$)
where the $\lambda\!\hsm =\!\hsm 0,\pm 1\,$
components arise from absorbing
a scalar Goldstone boson $h_n^{55}$
($\lambda\!\!=\!\!0$) and a vector Goldstone boson $h_n^{\mu 5}$
($\lambda\!\!=\!\!\pm 1$) in the 5d graviton field.\
We demonstrate that
{\it the GRET provides a general energy-cancellation mechanism}
guaranteeing that the leading energy dependence of $N$-particle longitudinal KK graviton amplitudes
($\propto\! E^{2(N+1+L)}$)
must cancel down to a much lower energy power
($\propto\! E^{2(1+L)}$)
by an energy factor of $E^{2N}$,
as enforced by matching the energy dependence of the corresponding leading gravitational KK Goldstone amplitudes,
where $L$ denotes the loop number of the relevant Feynman diagram.\ 
For the four-point longitudinal KK graviton scattering
amplitudes at tree level, this proves the energy cancellations
$\,E^{10}\!\ito E^2\,$, which explains the result of the
recent explicit calculations of 4-longitudinal KK graviton amplitudes\,\cite{Chivukula:2019S}\cite{Chivukula:2019L}\cite{Kurt}.
\\[1.4mm]
\hspace*{1em}%
The double-copy approach has profound importance 
for understanding the quantum gravity
because it uncovers the deep gauge-gravity connection
at the scattering $S$-matrix level,
$\rm{GR} \!=\! (\rm{Gauge}$
$\rm{Theory})^2$\,\cite{Elvang:2013}.\
The conventional double-copy method with
color-kinematics (CK) duality of
Bern-Carrasco-Johansson (BCJ)\,\cite{BCJ:2008}\cite{BCJ:2019}
was proposed to connect scattering amplitudes between 
the massless Yang-Mills (YM) gauge theories 
and the massless GR theories.\
It was inspired by
the Kawai-Lewellen-Tye (KLT) relation\,\cite{KLT}
which connects the product of two scattering amplitudes of
open strings to that of the closed string
at tree level\,\cite{Tye-2010}.
\\[1mm]
\hspace*{1em}%
Extending the conventional double-copy approach,
we construct the massive KK graviton (Goldstone) amplitudes
from the massive KK YM gauge (Goldstone) amplitudes
under high energy expansion at
the leading order (LO) and at the next-to-leading order (NLO).
This provides an extremely efficient way
to derive the complicated massive KK graviton amplitudes
from the massive KK gauge boson amplitudes, and gives a deep 
understanding on the structure of the KK graviton amplitudes.
\\[1mm]
\hspace*{1em}%
Because the LO amplitudes of the longitudinal KK gauge bosons
and of their KK Goldstone bosons have $\mO(E^0M_n^0)$
and are equal (leading to the KK GAET)\,\cite{5DYM2002},
our double-copy approach demonstrates
that the reconstructed LO amplitudes of the
longitudinal KK gravitons and of the KK Goldstone bosons have $\mO(E^2M_n^0)$, and must be equal to each other 
(leading to the KK GRET),
where $\Mn$ denotes the relevant KK mass.
Our double-copy construction further proves that the residual
term of the GRET belongs to the NLO, which has
$\mO(E^0M_n^2)$ and is suppressed relative to the LO KK
Goldstone boson amplitude of $\mO(E^2M_n^0)$.\
Finally, we further construct an exact double-copy of the
KK graviton scattering amplitudes at the NLO. 

\vspace*{-2mm}
\section{\hspace*{-2.5mm}\boldmath{$R_\xi^{}$}\,Gauge-Fixing~and~Geometric\,Higgs\,Mechanism}
\label{sec:2}
\vspace*{-1mm}


We consider the compactified GR5 under the orbifold $S^1\!/\ZZ$
where the fifth dimension is a line segment
$0 \!\leqslant x^5$ $\leqslant\!\hsm L\,(=\!\pi r_c)$,
with $r_c^{}$ being the compactification radius.
Extension to the case of warped 5d space\,\cite{RS} does not cause 
conceptual change regarding our current study.
Thus, the 5d Einstein-Hilbert (EH) action
takes the following form:
\begin{equation}
\label{eq:LagEH}
\hspace*{-5mm}
S_{\rm{EH}}^{} \,= \int\!\! \td^5 x \, \hat\La_{\rm{EH}}
\,= \int \!\! \td^5 x \frac{2}{\,\hka^2\,}
\sqrt{-\hat{g}\,} \hat{R} \,,
\end{equation}
where the coupling constant
$\,\hka = \!\sqrt{32 \pi \hat{G}\,}\,$.

\vspace*{1mm}

Then, we expand the 5d EH action \eqref{eq:LagEH}
under the metric perturbation
$\,\hg _{AB}^{} \!= \! \heta_{AB}^{} \!+\! \hka \hs \hh_{AB}^{}$\,,
where 
$\,\heta^{}_{AB}\!=\diag$ $(-1,1,1,1,1)$\,
is the 5d Minkowski metric.
Thus, we can express the 5d graviton field $\,\hh_{AB}^{}$ 
as follows:
\begin{equation}
\label{eq:hDecom}
\hh_{AB} \,=\,
\begin{pmatrix}
\hh_{\mn} \!-\! \frac{1}{2}\eta_{\mn} \hat{\phi} ~&~ \ \hh_{\mu5}
\\[1.5mm]
\hh_{5\nu} &\ \hat{\phi}
\end{pmatrix} \!.
\end{equation}
\\[-2mm]
Under the compactification of $S^1\!/\ZZ$\,,
the spin-2 field $\hh_{\mn}$
and scalar field $\phih\,(\equiv \hh^{55})$
are $\ZZ$ even, while the vector field
$\hat{\A}_\mu^{}\,(\equiv \hh_{\mu 5}^{})$
is $\ZZ$ odd.\
After compactification, we derive the 
4d effective Lagrangian 
for both the zero-modes and KK-modes
($h_n^{\mu\nu}\hsm ,\, \A_n^{\mu}\hs ,\, \phi_n^{}$)\,\cite{supp}.\

\vspace*{1mm}

We further construct a general
$R_\xi^{}$-type gauge-fixing term as follows:
\\[-5mm]
\begin{equation}
\label{eq:GF}
\La_{\rm{GF}}^{} \,=\,
 -\sum_{n=0}^{\infty}\frac{1}{\xi_n}\!
\[\! (F_n^\mu)^2 + (F_n^5)^2  \!\] ,
\end{equation}
\\[-4mm]
where the gauge-fixing functions $\hs (F_n^\mu ,\,F_n^5)\hs$ 
take the following form\,\cite{Hang:2021fmp},
\\[-5mm]
\beqs
\label{eq:GF2}
\begin{align}
F_n^\mu &\,=\, \pd_\nu h^{\mn}_n \!-\!
\(\!1 \!-\! \fr{1}{\,2\xi_n}\)\!
\pd^\mu h_n^{} +\xi_n^{}M_n^{}\An \,,
\\[.5mm]
F_n^5 &\,=\, \fr{1}{2} \!\!\( M_n h_n \!-\!
3 \xi_n M_n \phin\! + 2 \pd_{\mu} \A_n^{\mu} \) \!.
\end{align}
\eeqs
The above $R_\xi^{}$ gauge-fixing can ensure the kinetic terms
and propagators of the KK fields
$(h_n^{\mn}\!,\,\A_n^\mu,\,\phin)$
to be diagonal.\
In the limit of $\,\xi_n^{}\!\ito\infty\hs$,$\hs$
we recover the unitary gauge where the KK Goldstone bosons
$(\A_n^\mu,\,\phin)$ are fully absorbed (eaten) by
the corresponding KK gravitons $h_n^{\mn}$ 
at each KK level-$n$\,.\ 
This realizes a Geometric Gravitational ``Higgs'' Mechanism 
for KK graviton mass-generations. 

\vspace*{1mm}

Then, we derive the propagators of KK gravitons and KK Goldstone bosons under the $R_\xi^{}$ gauge-fixing \eqref{eq:GF} \cite{supp}.
For Feynman-'t\,Hooft 
gauge ($\xi_n^{} \hsmx\!=\! 1$), 
the KK propagators take the following simple forms:
\beqs
\label{eq:D-xi=1}
\begin{align}
\hspace*{-3mm}
\D_{nm}^{\mn\ab}(p)
&= -\frac{\,\ii\dnm\,}{2}
\frac{\,\eta^{\mu \al}\eta^{\nu \be}\!+\!\eta^{\mu \be}\eta^{\nu\al}\!-\!
\eta^{\mu\nu}\eta^{\al\be}\,}
{\,p^{2}\hsm +\hsm M_{n}^{2}\,} \,, 	
\label{eq:Dhh}
\\
\hspace*{-3mm}
\D^{\mn}_{nm}(p)  &=
-\frac{\,\ii\eta^{\mn}\dnm\,}
{\,p^{2}\hsm +\hsm M_{n}^{2}\,} \,,  \quad
\D_{nm}^{} (p) =
-\frac{\,\ii \dnm\,}{\,p^2 \hsm +\hsm M_n^2\,} ,
\label{eq:DAA-D55}
\end{align}
\eeqs
which all share the same mass-pole $p^2\!\hsm =\!-\Mnn$\,.

\vspace*{1mm}

Strikingly, we observe that
our massive KK graviton propagator \eqref{eq:Dhh}
has a smooth limit for $\!M_n \!\ito 0$\,,
under which Eq.\eqref{eq:Dhh} reduces to the conventional
massless graviton propagator of Einstein gravity.
Hence, we have proven that {\it the KK graviton propagator
is free from the vDVZ discontinuity}\,\cite{vDVZ}
which is a longstanding problem plaguing 
the conventional Fierz-Pauli massive gravity theory 
and alike\,\cite{PF}\cite{Hinterbichler:2012}.\ 
This is because the GHM under KK compactification guarantees that
the physical degrees of freedom of each KK graviton 
are {\it conserved before and after taking the massless
limit $\!M_n \!\ito 0$,} i.e., $5=2+2+1\hs$.
This demonstrates that the compactified KK GR theory can 
consistently realize the mass-generation for spin-2 KK gravitons.\  
We can also derive the unitary-gauge propagator of 
KK gravitons by taking the limit 
$\,\xi_n^{}\!\!\to\infty$ \cite{Hang:2021fmp}.

\vspace*{-2mm}
\section{\hspace*{-2.5mm}GRET Formulation for the GHM}
\label{sec:3}
\vspace*{-1mm}


In the previous section, we analyzed the geometric Higgs mechanism 
at the Lagrangian level. 
In this section, we further formulate the GRET,
which {\it realizes the geometric gravitational Higgs mechanism 
at the $S$-matrix level.} 
Using the gauge-fixing terms \eqref{eq:GF}-\eqref{eq:GF2} and following
the method of Ref.\,\cite{ET94}, we derive a Slavnov-Taylor-type identity in the momentum space:
\begin{equation}
\label{eq:F-identityP}
\hspace*{-1mm}
\la 0| F_{n_1}^{\mu_1}(k_1^{})F_{n_2}^{\mu_2}(k_2^{})\cdots
F_{m_1}^{5}(p_1^{})F_{m_2}^{5}(p_2^{})\cdots \Phi
|0 \ra
= 0\,,
\end{equation}
where $\Phi$ denotes any other on-shell physical fields 
after the Lehmann-Symanzik-Zimmermann (LSZ) amputation 
and each external momentum obeys the on-shell condition
$\,k_j^2\!=\!-M_{n_{\hsm j}^{}}^2$ or\,
$\,p_j^2\!=\!-M_{m_{\hsm j}^{}}^2$.\ 
The identity \eqref{eq:F-identityP} is a direct consequence of the
diffeomorphism (gauge) invariance of the compactified KK 
theory\,\cite{Hang:2021fmp}\cite{ET94}.

\vspace*{1mm}

Under the Feynman-'t\,Hooft gauge ($\xi_n\!=\!1$) and at the tree level,
we can directly amputate each external state by multiplying
the propagator-inverse $(k^2\!+\!M_n^2)\ito 0$\,
for Eq.\eqref{eq:F-identityP}.
Thus, we derive\,\cite{Hang:2021fmp} 
the following GRET identity 
which connects the longitudinal KK graviton amplitude 
to the corresponding KK Goldstone
amplitude plus a residual term:
\beqs
\label{eq:GET}
\begin{align}
\label{eq:GET1}
&\hspace*{-1.5mm}
 \M [h^L_{n_1^{}}\!,\cdots\!,h^L_{n_{\hsm N}^{}},\Phi]
\,=\,
\M [\phi_{n_1^{}}^{}\!,\cdots\!,\phi_{n_{\hsm N}^{}}^{} \hs ,\Phi]
+\M_{\!\Delta}^{}\hs ,
\\[0.8mm]
\label{eq:GET-RT}
&\hspace*{-1.5mm}
\M_{\!\Delta}^{} =\!
\sum_{1\leqq k \leqq N}\!\!\!
\M [\{\widetilde{\Delta}_{n_{\hsm k}}^{},\phin\},\Phi] \,,
\\[-8mm]
\nonumber
\end{align}
\eeqs
where  $\widetilde{\Delta}_n^{} \!= \vnt \!-\thn^{}$,\, $\vnt\!=\!\vt_{\mn}^{}h_n^{\mn}$,\, and
$\thn^{}\!\!=\!\!\sqrt{2/3}\,\eta_{\mn}^{}h_n^{\mn}$.
The tensor
$\,\vt^{\mn}\!\!=\!\vep_L^{\mn}\!\!-\!\sqrt{2/3\,}\,
\vep_S^{\mn}\!\! =\mO(E^0)$,\,
and $\,(\vep_L^{\mn}\hsm ,\,\vep_S^{\mn})$
are the (longitudinal,~scalar) polarization tensors
of the KK graviton $h_n^{\mn}$.\
We can extend the GRET \eqref{eq:GET} up to loop levels
and valid for all $R_\xi^{}$ gauges
by using the gravitational BRST identities\,\cite{GET-2},
similar to the ET formulation
in the 5d KK YM theories\,\cite{KK-ET-He2004}
and in the 4d standard model (SM)\,\cite{ET94,ET96,ET-Rev}.\ 

\vspace*{1mm}

Inspecting the scattering amplitudes in the GRET identity 
\eqref{eq:GET1},
we can make direct power counting 
on the leading $E$-dependence of individual
Feynman diagrams for each amplitude.\
For the four-particle scattering, 
the longitudinal KK graviton amplitude 
on the left-hand-side 
of \eqrefe{eq:GET1}
contains individual contributions via
quartic interactions or via exchanging KK-mode
(zero-mode) gravitons.\ Since each external longitudinal
KK graviton has polarization tensor
$\,\vep_L^{\mn} \!\!\supset\! k^\mu k^\nu\!/\Mnn\hs$,
the leading individual contributions behave as
$\mO(E^{10})\hs$.\
But we observe that on the right-hand-side (RHS) 
of \eqrefe{eq:GET1}, the external states
in all amplitudes have no superficial enhancement
or suppression factor.\ 
Thus, by power counting on the KK amplitudes,  
we find that the RHS of \eqrefe{eq:GET1}
(including the residual term 
$\M_{\!\Delta}^{}$) scales as $\mO(E^2)\hs$. 
Hence, the GRET identity \eqref{eq:GET} provides 
{\it a general mechanism} for the large energy cancellations
of $E^{10} \!\!\ito E^2$ 
in the four-longitudinal KK graviton amplitudes.

\vspace*{1mm}

We have further developed 
a generalized energy-power counting method\,\cite{supp}
for the massive KK gauge and gravity theories,\ 
by extending the conventional 4d power counting rule of 
Steven Weinberg for the nonlinear sigma model of 
low energy QCD\,\cite{weinberg}\cite{steve-foot}.\
With this and the GRET \eqref{eq:GET},
we can prove a general energy cancellation 
$E^{2(N+1+L)}\!\ito\! E^{2(1+L)}$
in the $N$-point longitudinal KK graviton amplitudes,
which cancels the leading energy-dependence by
$E^{2N}\!$ powers\,\cite{supp}.\
For $N$-point longitudinal KK gauge boson amplitudes,
we also prove\,\cite{Hang:2021fmp} a general energy cancellation 
of $\,E^4\!\ito\hsm E^{4-N-\delta},\hs$ 
which cancels the leading energy powers by
$\,E^{N+\delta}$,
with $\,\delta \!=\![1\hsm -\hsm (-1)^N]/2\,$.
We will establish {\it a new correspondence between the
two types of energy cancellations} 
in the $N$-point KK gauge boson scattering amplitudes and 
KK graviton scattering amplitudes in Sec.\ref{sec:5}.

\vspace*{-2mm}
\section{\hspace*{-2.5mm}KK Graviton Scattering\,Amplitudes from GRET}
\label{sec:4}
\vspace*{-1mm}

\begin{figure}[t]
\centering
\includegraphics[width=8.5cm]{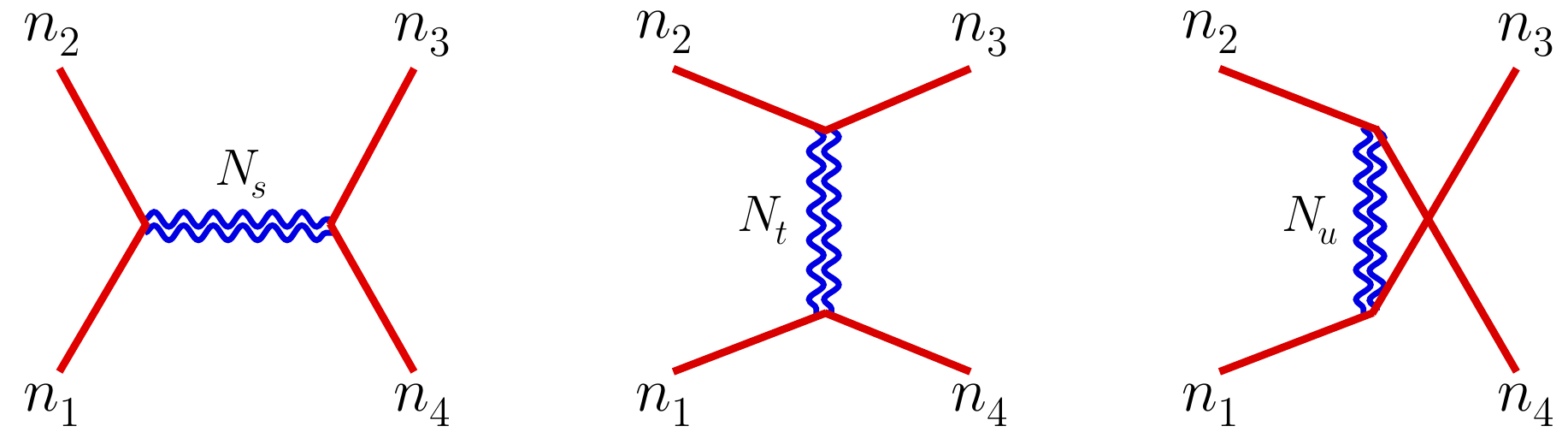}
\vspace*{-2mm}
\caption{Feynman diagrams for the scattering of 
gravitational KK Goldstone bosons, 
$\phi_{n_1}^{}\phi_{n_2}^{}\!\!\ito \phi_{n_3}^{}\phi_{n_4}^{}$,
by exchanging the KK gravitons of level-$N_j$
($\,j=s,t,u)$ at the tree level, which contribute 
the leading energy-dependence of $\mO(E^2)$.}
\label{fig:nn-nn-h}
\vspace*{-2mm}
\end{figure}

In the following,
we demonstrate explicitly how the GRET holds.\
For this purpose, we compute the gravitational 
KK Goldstone boson scattering 
amplitude
$\MT [\phi_{n_1}^{}\phi_{n_2}^{} \!\!\ito 
\phi_{n_3}^{}\phi_{n_4}^{}]$
($n_j^{}\!\!\geqq\!\! 1$).\   
The relevant Feynman diagrams having leading energy contributions 
are shown in Fig.\,\ref{fig:nn-nn-h}.

\vspace*{1mm}

For the elastic scattering, we set the KK numbers of all external
states as $\,n_i^{} \!\!=\! n\,$
and of internal states as
$\,N_j \hsm\!=\! 0,\hs 2n\,$.\  
Then, summing up the contributions of Fig.\,\ref{fig:nn-nn-h}
and making high energy expansion,
we derive the following LO scattering amplitude of 
the gravitational KK Goldstone bosons:
\begin{equation}
\vspace*{-1.5mm}
\label{eq:AmpExHLO}
\hspace*{-5mm}
\MT_0^{} \,=\,
\frac{\,3\ka^2\,}{\,128\,}
\frac{~(7\!+ \cos 2\theta)^2\,}{\sin^2\!\theta}\,
s \,,
\vspace*{-1.mm}
\end{equation}
where
$\,\MT_0^{}\hsm =\hsm\MT_0^{}[\phin\phin\!\ito \phin\phin]\hs$.\
To compare our Eq.\eqref{eq:AmpExHLO} with the
corresponding longitudinal KK graviton amplitude of
Refs.\,\cite{Chivukula:2019S}\cite{Chivukula:2019L}, we rescale
our coupling $\,\ka\!\ito\ka/\!\sqrt{2\,}$ to match their
normalization and find that
{\it the two amplitudes are equal at the LO\,}:
\beq
\label{eq:GRET-LO-4n}
\M_0^{}[h^n_L h^n_L \ito h^n_L h^n_L] \,=\,
\MT_0^{}[\phin\phin\hsm\ito \phin\phin]\hs . 
\eeq 
Namely, 
$\M_0^{}\!=\!\MT_0^{}\hs$,
where we denote 
$\,\M_0^{}\!=\!\M_0^{}[h^n_L h^n_L \!\ito\! h^n_L h^n_L]\,$
and
$\,\MT_0^{}\!=\!\MT_0^{}[\phin\phin\hsm\ito \phin\phin]\,$.
From the GRET identity \eqref{eq:GET1}
[and \eqrefe{AmpK-hLh5-nnnn}], this means that the
residual term \eqref{eq:GET-RT} belongs to the NLO:
\beq
\M_{\!\Delta}^{}\!=\M -\MT = 
\dM\!-\dMT =\mO(E^0\Mnn)\hs .
\eeq 
%
and thus is much smaller.\ 
We have further computed the exact tree-level Goldstone boson amplitude
$\MT\hs$ by including all the subleading diagrams\,\cite{supp}.

\vspace*{1mm}

For inelastic scattering of gravitational KK Goldstone bosons,
we compute the four-point amplitudes and 
find that the LO inelastic amplitude
is connected to the LO elastic amplitude \eqref{eq:AmpExHLO} by
the following relation: 
\beqa 
\MT[\phi_{n_1}\phi_{n_2}\!\!\ito\phi_{n_3}\phi_{n_4}]
\,=\, \zeta \MT[\phi_n\phi_n \!\ito\! \phi_n\phi_n]\hs ,
\eeqa 
%
where $\,\zeta\!=\! 2/3\,$
for $\,n_1^{}\!=\!n_2^{}\!\neq\! n_3^{}\!=\!n_4^{}\,$,\,
and $\,\zeta\!=\!1/3\,$
for the cases with KK numbers
$(n_1^{},n_2^{},$ $n_3^{},n_4^{})$ having
no more than one equality.
{\linespread{1.85}
\begin{table*}[t]
\centering
\caption{%
\baselineskip 13pt
Kinematic numerators of the LO and NLO scattering amplitudes
for KK longitudinal gauge bosons and KK Goldstones
as defined in Eq.\eqref{Amp-ALA5-nnnn},
where $(\NN_j,\,\NNt_j) = (\NN_j^0,\,\NNt_j^0) \!+\! (\da\NN_j,\,\da\NNt_j) = \mO(E^2M_n^0)+\mO(E^0\Mnn)$,
and $(s_{\theta}^{},\,\ct)=(\sin\hsm\theta,\,\cos\hsm\theta)$.}
\vspace*{1mm}
\begin{tabular}{c||c|c|c||c|c|c||c|c|c}
\hline\hline
Numerators \hs
&\hsx\hsx $\NN_s^{}$ \hsx
&$\NN_t^{}$
&$\NN_u^{}$
&$\quad\NNt_s^{}\quad$
&$\NNt_t^{}$
&$\NNt_u^{}$
&$\,\NN_s^{}\!-\!\NNt_s^{}\,$
&$\,\NN_t^{}\!-\!\NNt_t^{}\,$
&$\,\NN_u^{}\!-\!\NNt_u^{}\,$ \\
\hline
$\NN^0_{\hsm j}/s_{\hsm j}$
&\large\hsx\hsx$\frac{\,5\ct\,}{2}$\hsx
&\large$\frac{\,13+5 \ct + 4 \ctt\,}{2(1+\ct)\,}$
&\large$ -\frac{\,13 -5 \ct + 4 \ctt\,}{2(1-\ct)\,}$
&\large$-\frac{\,3\ct\,}{2} $
&\large$\frac{\,3(3-\ct)\,}{2(1+\ct)\,}$
&\large$-\frac{\,3(3+\ct)\,}{2(1-\ct)\,}$
&$4 \ct$
&$4 \ct$
&$4 \ct$
\\ \hline
$\dNN_{\hsm j}^{}/\Mnn$
&\hsx\hsx$4\ct$\hsx
&\large$\frac{\,2 (2-3\ct-2\ctt-\cttt)\,}{1+\ct\,}$
&\large$-\frac{\,2 (2+3\ct-2\ctt+\cttt)\,}{1-\ct\,}$
&$\,4\ct\,$
&\large$-\frac{\,8\ct\,}{\,1+\ct\,}$
&\large$-\frac{\,8\ct\,}{\,1-\ct\,}$
&$0$
&$8 s_{\theta}^2$
&$-8 s_{\theta}^2$ \\
\hline\hline
\end{tabular}
\label{NumeratorTable}
\label{tab:1}
\end{table*}}

\vspace*{-3mm}
\section{\hspace*{-2.5mm}Double-Copy Construction of 
Massive\\ KK Scattering Amplitudes}
\label{sec:5}
\vspace*{-2mm}


The double-copy construction for the massive KK gauge/gravity scattering amplitudes are highly nontrivial.\
We make the first serious attempt for an explicit double-copy construction of KK amplitudes under high energy expansion.\
We present the four-point elastic scattering 
amplitudes of longitudinal KK gauge bosons (Goldstones)
at the LO and NLO:
\beqs
\label{Amp-ALA5-nnnn}
\begin{align}
\label{Amp-AL-nnnn}
\hspace*{-3.mm}
\TT &=  \sum_{j}  \!\frac{\,g^2\CC_j \NN_{\hsm j}}{s_{\hsm j}^{}}
= \sum_{j} \!\frac{\,g^2\CC_j (\NN^0_{\hsm j} 
\!+ \da\NN_{\hsm j})}{s_{\hsm j}^{}}
= \TT_0^{} \!+ \dT ,
\\[-1mm]
\label{Amp-A5-nnnn}
\hspace*{-3mm}
\tT &=  \sum_{j}\!\frac{\,g^2\CC_j \NNt_{\hsm j}}{s_{\hsm j}^{}}
= \sum_{j}\!\frac{\,g^2\CC_j (\NNt^0_{\hsm j} \!+ \da\NNt_{\hsm j})}{s_{\hsm j}^{}}
= \tT_0^{} \!+ \da\tT ,
\end{align}
\eeqs
\\[-2mm]
where we have denoted $\,\T \!\hsm\equiv\!\hsm 
\T[A^{an}_L A^{bn}_L \!\ito\! A^{cn}_L A^{dn}_L]$ and
$\tT \!\equiv
\tT[A^{an}_5 A^{bn}_5 \!\ito\! A^{cn}_5 A^{dn}_5]$.\ 
We also define the SU($N$) color factors as
$(\CC_s,\, \CC_t,\, \CC_u ) \!\equiv\!
\(C^{abe}C^{cde}\!,\, C^{ade}C^{bce}\!,\,C^{ace}C^{dbe}\)$,
which obey the Jacobi identity
$\,\CC_s \!+ \CC_t \!+ \CC_u \!=\hsm 0\hs$.

\vspace*{1mm}

We present in Table\,\ref{tab:1}
the numerator factors ($\NN_{\hsm j},\,\NNt_{\hsm j}$) of
Eqs.\eqref{Amp-AL-nnnn}-\eqref{Amp-A5-nnnn}.\
Table\,\ref{tab:1} shows that
$(\T_0^{},\,\tT_0^{})\!\!=\mO(E^0M_n^0)$ and
$(\dT,\,\da\tT)=\mO(\Mnn /E^2)\hs$.\
We find that the sum of each set of  
the LO, NLO, and NNLO numerators 
of the KK gauge (Goldstone) scattering amplitudes 
in \eqrefe{Amp-ALA5-nnnn} 
violate the kinematic Jacobi identity 
by terms of $\mO(\Mnn)$ and $\mO(M_n^4/E^2)$,
respectively: 
\\[-4mm]
\beqs
\label{eq:sum-Nj-tNj}
\begin{align}
\label{eq:sum-Nj-LO}
\hspace*{-3mm}
&\sum_j\!\NN_j^0 = 10\hs\ct M_n^2  \,,~~~~
\sum_j\!\NNt_j^0 =\hsm -6\hs\ct M_n^2  \,,
\\
\label{eq:sum-dNj-NLO}
\hspace*{-3mm}
& \sum_j\!\da_1^{}\NN_j\hsm 
= \sum_j\!\da_1^{}\NNt_j \hsm =\hsm 
-2\hs (7\hsm +\hsm \ctt)\ct\csc^2\!\theta
\hs\Mnn \,,
\\
\label{eq:sum-dNj-NNLO}
\hspace*{-3mm}
& \sum_j\!\da_2^{}\NN_j\hsm 
=8\hs (31\hsm +\hsm \ctf)\hs\ct\csc^4\!\theta\hs M_n^4/s\,,
\\
& \sum_j\!\da_2^{}\NNt_j \hsm =\hsm 
32\hs (7\hsm +\hsm \ctt)\hs\ct\csc^4\!\theta\hs M_n^4/s\,,
\end{align}
\eeqs
\\[-5mm]
where $\,c_{n\theta}^{}\!=\hsm\cos(n\theta)\hs$,  
$\da\NN_j\!=\! 
\da_1^{}\NN_j\!+\da_2^{}\NN_j$, and
$\,\da\NNt_j\!=$ $\da_1^{}\NNt_j\!+\hsm \da_2^{}\NNt_j\hs$.\  
Hence, we cannot naively apply color-kinematics duality
for BCJ-type double-copy construction without making 
further modifications on these numerators. 

\vspace*{1mm} 

Inspecting the scattering amplitudes 
in Eq.\eqref{Amp-ALA5-nnnn},
we first observe that they 
are invariant under the following generalized gauge transformations 
of their numerators:
%
\beqa
\label{eq:GGtransf}
\NN_{\hsm j}^{\pp} = \NN_{\hsm j}^{} + s_{\hsm j}^{}\hs\Delta \,,
~~~~
\NNt_{\hsm j}^{\pp} = \NNt_{\hsm j}^{} + s_{\hsm j}^{}\hs
\widetilde{\Delta} \,.
\eeqa
%
We can determine the gauge-parameters 
$(\Delta,\,\widetilde{\Delta})$ 
by requiring the gauge-transformed
numerators to obey the Jacobi identities
$\sum_j\hsm\NN_{\hsm j}^{\pp}\hsm\!=\hsm 0\hs$ and
$\sum_j\hsm\NNt_{\hsm j}^{\pp}\hsm\!=\hsm 0\,$.\
Thus, we derive the following general solutions: 
\\[-3.5mm]
\beqa
\label{eq:sol-Delta-tDelta}  
\Delta =
-\frac{1}{\,4\Mnn\,}\!\sum_j \NN_j^{}\,,~~~~
\widetilde{\Delta}=
-\frac{1}{\,4\Mnn\,}\!\sum_j \NNt_j^{}\,,~~~~~
\eeqa
%
\\[-3.3mm]
which realize the BCJ-respecting numerators
$(\NN_j^{\pp},\,\NNt_j^{\pp})$.
Making high energy expansions on both sides of 
Eq.\ \eqref{eq:sol-Delta-tDelta}, we derive the expressions
of the gauge-parameters 
$(\Delta,\,\widetilde{\Delta})=
(\Delta_0^{}\hsm +\hsm\Delta_1^{},\,
 \widetilde{\Delta}_0^{}\hsm +\hsm\widetilde{\Delta}_1^{})\, 
$ at the LO and NLO$\hs$: 
%
\begin{equation}
\begin{aligned}
\Delta_0^{} &=
\fr{1}{4}(9\hsm +\hsm 7\hs\ctt)\hs\ct\csc^2\!\theta \,,~~~
\\
\widetilde{\Delta}_0^{} &=
\fr{1}{4}(17\!-\hsm\ctt)\hs\ct\csc^2\!\theta \,,
\\
\Delta_1 &=
-2(31\hsm +\hsm \ctf)\ct\csc^4\!\theta\,(\Mnn/s) \,, 
\qquad
\label{eq:sol-Delta01}
\\
\widetilde{\Delta}_1  &=
-8(7\hsm +\hsm \ctt)\ct\csc^4\!\theta\,(\Mnn/s) \,.
\end{aligned}
\end{equation}
%
With these, we further compute the new numerators
$(\NN_j^{\pp},\,\NNt_j^{\pp})$, and 
derive explicitly the LO results in Eq.\eqref{eq:N0j'}
and the NLO results in the Supplemental Material\,\cite{supp}.

\vspace*{1mm}

For the 5d KK YM (YM5) and 5d KK GR (GR5) theories,\
we expect the double-copy correspondence between the
KK gauge fields and KK graviton fields:
\beqa
A_n^{a\mu} \!\otimes\! A_n^{a\nu} &\longrightarrow&~ h^{\mn}_n\,,
\nn\\
A_n^{a5} \!\otimes\! A_n^{a5} &\longrightarrow&~ h^{55}_n \,,
\\
A_n^{a\mu} \!\otimes\! A_n^{a5} &\longrightarrow&~ h^{\mu 5}_n \,.
\nn 
\eeqa 
%
The physical spin-2 KK graviton field
$h^{\mn}_n$ arises from two copies of spin-1 KK gauge fields.\ 
The KK Goldstone boson $A_n^{a5}$ of the YM5 
has its double-copy counterparts $h^{55}_n\,(=\!\!\phin)$ and
$h^{\mu 5}_n\,(=\!\!\A^\mu_n)$
which correspond to the scalar and vector KK Goldstone bosons
in the compactified GR5 theory.\ 
The double-copy correspondence between the 
longitudinal KK modes,
$A_L^{an} \!\!\otimes\! A_L^{an}\!\!\to\!\! h_L^n$,\,
is {\it highly nontrivial even at the LO} of high energy expansion,
because $(A_L^{an},\hs h_L^n)$ do not exist in 
$\Mn\!\ito 0$ limit and the KK Goldstone bosons
$(A_5^{an},\hs \phin)$ become physical states  
in massless limit.\ 
Hence, this double-copy is consistently realized
only because we can use the KK GRET (GAET) to connect
$h_L^n\,(A_L^{an}$)
amplitudes to the $\phin\,(A_5^{an}$) amplitudes
under the $\Mn/E\ito\hs 0$\, limit where we can hold the 
KK mass $\Mn$ fixed and take the energy $\hs E\ito\infty\hs$.  

\vspace*{1mm}

Then, we extend the conventional double-copy
method \cite{BCJ:2008}\cite{BCJ:2019} to the massive
KK YM theory {\it under high energy expansion.}\
We apply the correspondence of 
color-kinematics duality 
$\,\CC_j\!\ito \NN'_{\hsm j}\,$ to Eq.\eqref{Amp-AL-nnnn} 
and 
$\,\CC_j\!\ito\NNt'_{\hsm j}$ to Eq.\eqref{Amp-A5-nnnn}.\
Thus, we can construct the following four-particle 
KK graviton (Goldstone) amplitudes:
\beqs
\label{AmpK-hLh5-nnnn}
\begin{align}
\label{Amp-hL-nnnn}
\hspace*{-3mm}
\M &\hs =\hs   
\sum_{j}\!  
\frac{~c_0^{}\hs g^2 (\NN^{0\hs\pp}_{\hsm j} \!+   
\da\NN_{\hsm j}^{\pp})^2\,}{s_{\hsm j}^{}}
=\, \M_0^{}\! + \da\M \,,
\\[-1mm]
\label{Amp-phi-nnnn}
\hspace*{-4mm}
\MT &\hs =\hs   
\sum_{j}\! \frac{~c_0^{}\hs g^2 
(\NNt^{0\hs\pp}_{\hsm j} \!+
\da\NNt_{\hsm j}^\pp)^2\,}{s_{\hsm j}^{}}
=\hs \MT_0^{} \hsm + \da\MT \,,
\end{align}
\eeqs
%
where we have denoted the scattering amplitudes
$\M \!\equiv\! \M[h_L^nh_L^n \!\hsm\ito\!h_L^nh_L^n]$ 
and $\MT \!\equiv\! \MT[\pnnnn]$, 
and $c_0^{}$ is a conversion constant.

\vspace*{1mm}


From Table\,\ref{tab:1} and using 
Eqs.\eqref{eq:GGtransf}\eqref{eq:sol-Delta01}, 
we find that the LO numerators  
$(\NN_{\hsm j}^{0\hs\pp},\,\NNt_{\hsm j}^{0\hs\pp})$ 
are {\it mass-independent}  and {\it equal to each other$\,$}:
%
\beqs
\label{eq:N0j'}
\begin{align}
\NN_s^{0\hs\pp} &\!= \NNt_s^{0\hs\pp} \!= 
\frac{~s\hs (7\!+\!\ctt)\ct~}{2\sin^2\!\theta} \hs, 
\\ 
\NN_t^{0\hs\pp} &\!= \NNt_t^{0\hs\pp} \!=
-\frac{~s\hs (42\!-\!15\ct\!+\!6\ctt\!-\!\cttt)~}
{16\hs (1 \!-\! \ct)}
\hs ,\hspace*{5mm} 
\\ 
\NN_u^{0\hs\pp} &\!= \NNt_u^{0\hs\pp} \!= 
\frac{~s\hs (42 \!+\! 15\ct \!+\! 6\ctt \!+\! \cttt)~}
{16\hs (1 \!+\! \ct)} \hs .
\end{align}
\eeqs
\\[-2.5mm]
This demonstrates the {\it equivalence} 
between the two leading-order KK amplitudes
at $\mO(E^0 M_n^0)$, $\,\T_0^{}=\tT_0^{}\,$,  
which explicitly realizes the KK GAET. 
With these and using our LO double-copy formulas in Eq.\eqref{AmpK-hLh5-nnnn}, 
we can reconstruct the KK GRET:
\\[-5mm]
\beqa
\label{eq:ML0=M50}
\M_0^{}[\rm{DC}] \,=\, \MT_0^{}[\rm{DC}]\,,
\eeqa
which is of $\,\mO(E^2M_n^0)\hs$.\ 
We stress that as expected,\ these LO amplitudes are 
{\it mass-independent} and thus the LO double-copy 
can hold universally.\   We further find that
after setting the overall conversion constant of 
Eq.\eqref{AmpK-hLh5-nnnn} as
$\,c_0^{} \!=\!-\ka^2/(24g^2)\hs$,\,
the reconstructed LO KK amplitude
$\,\M_0^{}\,(\MT_0^{})\,$
just equals the LO KK Goldstone amplitude \eqref{eq:AmpExHLO} 
and the corresponding LO longitudinal KK graviton 
amplitude\,\cite{supp}.\ 
Hence, our double-copy prediction \eqref{eq:ML0=M50}
can prove (reconstruct) the KK GRET 
$\,\M_0^{}\!=\!\MT_0^{}$\, from the KK GAET 
$\,\T_0^{}\hsm =\hsm\tT_0^{}\,$.\
We derived this GRET relation in Eq.\eqref{eq:GRET-LO-4n}
by direct Feynman-diagram calculations.\
Note that the KK GAET relation 
$\,\T_0^{}\hsm =\hsm\tT_0^{}\,$ 
can hold for general $N$-point longitudinal KK gauge (Goldstone) 
amplitudes\,\cite{5DYM2002}\cite{KK-ET-He2004}.\ 
Hence, making double-copy on both sides of 
$\,\T_0^{}\!=\!\tT_0^{}\,$ 
can establish the GRET \eqref{eq:ML0=M50} to hold for
$N$-point longitudinal KK graviton (Goldstone) amplitudes.
From this, we can further establish 
{\it a new correspondence}
between the two types of energy cancellations 
in the $N$-longitudinal KK gauge boson amplitudes 
and in the corresponding $N$-longitudinal KK graviton amplitudes
(cf.\ the discussion around the end of Sec.\,\ref{sec:3}).

\vspace*{1mm}

Next, we use the double-copy formulas 
\eqref{Amp-hL-nnnn}-\eqref{Amp-phi-nnnn} to reconstruct 
the four-point longitudinal KK graviton amplitude and
the corresponding KK Goldstone boson amplitude 
at the NLO:
\beqs
\label{eq:dM-dmT-DC}
\begin{align}
\hspace*{-3mm}
\frac{\da\M \hsm (\rm{DC})\,}{\ka^2 \Mnn}\hsm &=
-\frac{\,5(1642 \hsm +\hsm  297\hs\ctt 
\hsm +\hsm 102\hs\ctf\hsm +\hsm 7\hs\cts)\,}
{768 \sin^4 \hsmx \theta}\hs ,
\\
\hspace*{-3mm}
\frac{\da\MT \hsm (\rm{DC})}{\ka^2\Mnn}\hsm &=
-\frac{\,6386 \hsm +\hsm 3837 \hs\ctt 
\hsm +\hsm 30\hs\ctf\hsm -\hsm 13\hs\cts\,}
{768 \sin^4 \hsmx \theta}\hs .
\end{align}
\eeqs
They have {\it the same size} of $\mO(\ka^2\Mnn)$ and 
{\it the same angular structure} of 
$\hs (1,\,\ctt,\,\ctf,\,\cts)\!\times\hsmx\csc^4\hsmx\theta\,$
as the original NLO amplitudes $(\da\M ,\, \da\MT)$
derived from Feynman diagram calculations\,\cite{supp},
though their numerical coefficients still differ.\
Then, using Eq.\eqref{eq:dM-dmT-DC} 
we compute the difference between the two
double-copied NLO amplitudes
$\,\Delta\M (\rm{DC})\!\!=\!\da\M \!-\! \da\MT\,$
and compare it with the NLO amplitude-difference
$\Delta\M (\rm{GR5})$
by Feynman diagram calculations 
in the KK GR5 theory:
\\[-5mm]
\beqs
\label{eq:Diff2-AmpGR5DC-nnnn}
\begin{align}
\label{eq:Diff2-AmpGR5-nnnn}
\hspace*{-2.5mm}
\Delta\M (\rm{GR5}) &=
-\fr{3}{2}\ka^2\!M_n^2
\(19.5 + \ctt\) ,
\\[1mm]
\label{eq:Diff2-AmpDC-nnnn}
\hspace*{-2.5mm}
\Delta\M (\rm{DC}) &=
-\ka^2 M_n^2\,( 7 \!+ \ctt )    \,.
\end{align}
\eeqs
\\[-5mm]
We find that they also have {\it the same size} of $\mO(\ka^2\Mnn)$ 
and {\it the same angular structure} of $\hs (1,\,\ctt)\hs$.\ 
Eq.\eqref{eq:Diff2-AmpGR5-nnnn} shows 
that the difference $\Delta\M (\text{GR5})$ 
between the original NLO amplitudes exhibits 
a striking precise cancellations of the angular structure
$\hs (1,\,\ctt,\,\ctf,\,\cts)\!\times\hsm\csc^4\hsmx\theta\,$
to $\hs (1,\,\ctt)\hs$. 
Impressively, {\it our double-copied NLO amplitude-difference
$\Delta\M (\text{DC})$ in Eq.\eqref{eq:Diff2-AmpDC-nnnn}
can also realize the same type of the precise angular cancellations.}

\vspace*{1mm}

The above extended NLO double-copy results 
\eqref{eq:dM-dmT-DC} and \eqref{eq:Diff2-AmpDC-nnnn}
are truly encouraging, 
because they already give {\it the correct structure}
of the NLO KK amplitudes including {\it the precise 
cancellations} of the angular dependence in
Eqs.\eqref{eq:dM-dmT-DC}-\eqref{eq:Diff2-AmpGR5DC-nnnn}.\
These strongly suggest that our massive KK double-copy approach 
{is on the right track.}\  
Its importance is twofold:\ 
(i).~In practice, for our proposed KK double-copy method
{\it under high energy expansion}, the LO double-copy 
construction is the most important part 
because it newly establishes
the GRET relation $\M_0^{} \!=\! \MT_0^{}$ 
[Eq.\ \eqref{eq:ML0=M50}] 
from the GAET relation
$\,\T_0^{}\!=\!\tT_0^{}\,$ [Eq.\eqref{eq:N0j'} and below],
as will be shown in Eq.\eqref{eq:KKET-GET}.\ 
The NLO KK graviton amplitudes are relevant 
only when we {\it estimate the size} of the residual term 
$\M_{\!\Delta}^{}$ of our GRET \eqref{eq:GET} 
and here we do not need
the precise form of $\M_{\!\Delta}^{}$ 
{\it except to justify its size 
$\M_{\!\Delta}^{}\!\!=\mO(E^0\Mnn)$} 
by the double-copy construction [cf.\ Eq.\eqref{eq:RTerm}].\
This proves that {\it the residual term
$\M_{\!\Delta}^{}\!$ does belong to the NLO amplitudes and is neligible
for our GRET formulation in the high energy limit.}\
Hence, {we do not need any precise NLO double-copy here.}\
(ii).~In general, our current KK double-copy approach as the
first serious attempt to construct the massive KK graviton 
amplitudes has given {strong motivation and important guideline} 
for a full resolution of the exact double-copy beyond the LO.
Our further study has found out the reasons for the 
minor mismatch between the numerical coefficients of the 
double-copied NLO amplitudes \eqref{eq:dM-dmT-DC} 
and that of the direct Feynman-diagram calculations.\ 
One reason is due to the double-pole
structure in the KK amplitudes 
(including exchanges of both the zero-mode and KK-modes)
beyond the conventional massless theories, so the additional 
KK mass-poles contribute to our mass-dependent NLO amplitudes
and cause a mismatch.\
Another reason is because the exact polarization tensor of the  
(helicity-zero) longitudinal KK graviton is given by
$\,\vep_{L}^{\mn} \hsm\!=\!\( 
\ep_{+}^{\mu} \ep_{-}^{\nu}\!+\ep_{-}^{\mu} \ep_{+}^{\nu}\!
+2 \ep_{L}^{\mu} \ep_{L}^{\nu}\)\hsm/\sqrt{6\,}$ \cite{supp}, 
which constains not only the longitudinal product 
$\ep_{L}^{\mu} \ep_{L}^{\nu}$, 
but also the transverse products 
$\,\ep_{+}^{\mu} \ep_{-}^{\nu}\!+\ep_{-}^{\mu} \ep_{+}^{\nu}\hs$. 
Thus, the other scattering amplitudes
containing possible transversely polarized external 
KK gauge boson states should be included 
for a full double-copy besides the
four-longitudinal KK gauge boson amplitude 
in Eq.\eqref{Amp-ALA5-nnnn}. 

\vspace*{1mm}

With these in minds, we have further used a first principle approach  
of the KK string theory in our recent work\,\cite{Li:2021yfk}
to derive the extended massive KLT-like relations 
between the product of the KK open string amplitudes 
and the KK closed string amplitude.\
In the field theory limit, we can derive the exact double-copy relations
between the product of the KK gauge boson amplitudes and the 
KK graviton amplitude at tree level\,\cite{Li:2021yfk}.\ 
In such exact double-copy relations all the relevant helicity indices
of the external KK gauge boson states are summed over to match
the corresponding polarization tensors of the external KK graviton
states.\ The double-pole structure is also avoided by first
making the 5d compactification under $S^1$ (without orbifold)
where the KK numbers ($\pm n\!=\!\pm 1,\pm 2,\pm 3,\cdots$) 
are strictly conserved and
the amplitudes are ensured to have single-pole structure.\ 
Then,\ we can define the $\ZZ$-even\,(odd) KK states as  
$\,|n_\pm^{}\rangle\hsm =\hsm (|\hsmx +\hsm n\rangle 
\pm |\hsmx -\hsm n\rangle)\hsm /\!\sqrt{2\,}$,
and derive the amplitudes under $S^1\! /\ZZ$ compactification
from the combinations of those amplitudes under 
the $S^1$ compactification\,\cite{Li:2021yfk}.\ 
Using this improved massive double-copy approach, we can
exactly reconstruct all the massive KK graviton 
amplitudes at tree level.
%
For the four longitudinal KK graviton amplitudes
under $S^1\! /\ZZ$, we derive the following 
(BCJ-type) exact massive double-copy formula:
\begin{equation}
\label{eq:MBCJ}
\hspace*{-1.5mm}
\M = -\frac{\ka^2}{64}
\sum_{\fP}\!\sum_{j}\!\sum_{\lam^{}_k,\lam'_k} 
\!\prod_k C_{\lam_k^{}\lam'_k}^{} 
\frac{\,N_{\hsm j}^\fP(\lam_k^{})N_{\hsm j}^\fP(\lam_k')\,}{D_{\hsm j}^{}} \,,
\end{equation}
where $\{N_{\hsm j}\}$ denote the kinematic numerators under the $S^1$ compactification and each external KK gauge boson has 3 helicity states 
(with $\lam_k^{}\!=\!\pm 1,L$ and $k\!=\!1,2,3,4$).
We use 
$\fP\!=\!\{n_1^{},n_2^{},n_3^{},n_4^{}\}$ 
to label each possible combination of the KK numbers for external 
gauge bosons, which obey the condition of KK number conservation
$\sum_{k=1}^{4}\! n_k^{}\!=\!0$\,. 
For the elastic KK scattering, we have
\begin{equation}
\begin{aligned}
\fP =&\ \{\pm n,\pm n,\mp n,\mp n\}, \hspace*{2mm}
\{\pm n,\mp n,\pm n,\mp n\},~~~
\\
&\ \{\pm n,\mp n,\mp n,\pm n\}.
\end{aligned}
\end{equation}
In Eq.\eqref{eq:MBCJ}, $C_{\lam_k^{}\lam'_k}^{}$ 
denotes the coefficients in the longitudinal polarization tensor
of $k$-th external KK graviton \cite{supp},
$\vep_{L,k}^{\mn}\!=\! \sum C_{\lam_k^{}\lam'_k}^{}
 \ep_{\lam_k}^\mu\ep_{\lam'_k}^\nu$, 
where $\lam_k^{},\lam'_k\!=\!\pm 1,L$ 
are the helicity indices for the $k$-th external gauge boson.\
The denominator of \eqrefe{eq:MBCJ} is defined as
$D_j^{}\!=\!s_j^{}\!-\!M_j^2$, where  
$\,s_j^{}\!\in\!\{s,t,u\}\,$ and 
$M_j^2\!\in\!\{M_{n_1+n_2}^2,M_{n_1+n_4}^2,M_{n_1+n_3}^2\}$.

\vspace*{1mm} 

Then, we make high energy expansion for the corresponding
elastic amplitude of KK gauge bosons (under $S^1$ compactification)
at the LO and NLO: 
\beqs
\begin{align}
\label{AmpS1-Annnn}
\TT & \,=\,  g^2\sum_{j}\frac{\,\CC_j N_{\hsm j}^{\fP}}{D_{\hsm j}^{}}
= g^2 \sum_{j} \!\frac{\,\CC_j (N^{0,\fP}_{\hsm j} \!+ \da N^{\fP}_{\hsm j})}{s_{\hsm j}^{}}
\hspace*{7mm}
\nn\\
& \,=~  \TT_0^{} + \dT \,,
\\[1mm]
N_{\hsm j}^{\fP} &\,=~
\frac{\,D_{\hsm j}^{}\,}{s_{\hsm j}^{}}\! 
\(\!{N_{\hsm j}^{0,\fP}}+\da N^{\fP}_{\hsm j}\)\!.
\end{align}
\eeqs
%
With this, we expand the exact double-copy formula of the
longitudinal KK graviton amplitude \eqref{eq:MBCJ} 
under the high energy expansion of $1/s$\,:
%
\begin{align}
\label{eq:MBCJ-LO+NLO}
\M &= -\frac{~\ka^2\,}{64}\,
\sum_{\fP}\sum_{j} \! \sum_{\lam^{}_k,\lam'_k} 
\prod_k C_{\lam_k^{}\lam'_k} 
\frac{\,D_{\hsm j}^{}\,}{s_{\hsm j}^2} 
\nn\\
& \hspace*{5mm}
\times\!
\[\!N_{\hsm j}^{0,\fP}(\lam^{}_k)\!+\da N_{\hsm j}^{\fP}(\lam^{}_k)\!\]\!\hsm
\[\!N_{\hsm j}^{0,\fP}(\lam'_k)\!+\da N_{\hsm j}^{\fP}(\lam'_k)\!\]
\hspace*{4mm}
\nn\\[1mm]
&= \,\M_0^{}+\dM\,.
\end{align}
%
It can be proven that the above double-copied LO amplitude $\M_0^{}$ 
is equivalent to the LO amplitude given in 
Eq.\eqref{Amp-hL-nnnn}\,\cite{Li:2021yfk}.\ 
We explicitly compute the above LO amplitude $\M_0^{}$
and find that $\M_0^{}$ just equals that of Eq.\eqref{eq:AmpExHLO}
as well as Eq.\eqref{eq:dM-dMT-GR5xx}\,\cite{supp}.
Then, we further compute the above double-copied NLO amplitude
$\dM$ as follows:
\beq 
\label{eq:DC-Amp-hL-NLOx}
\hspace*{-1.5mm}
\da\M 
= -\frac{\ka^2 M_n^2}{256}
(1810 + 93\hs\ctt +\hsm 126\hs\ctf +\hsm 
19\hs\cts)\hsm \csc^4 \hsmx\theta  \hs .
\eeq
We find that this fully agrees with the 
exact NLO elastic KK graviton amplitude 
derived from the direct Feynman diagram calculation 
in Eq.\eqref{eq:Amp-E-2hLxx} 
of the supplemental material\,\cite{supp}.
The above analysis is an explicit demonstration that
we can realize the exact (BCJ-type) massive double-copy construction
of the four-point KK graviton amplitudes
in Eq.\eqref{eq:MBCJ}, as well as
the precise double-copy of 
the KK graviton amplitudes
\eqref{eq:MBCJ-LO+NLO}-\eqref{eq:DC-Amp-hL-NLOx}
at both the LO and NLO of the high energy expansion. 
We will systematically pursue this new direction in our future work.

\vspace*{1mm}

Finally, it is very impressive
that our improved massive double-copy construction
of the longitudinal KK graviton (KK Goldstone) amplitude 
in Eq.\eqref{AmpK-hLh5-nnnn} is based on the pure longitudinal KK 
gauge (KK Goldstone) amplitude \eqref{Amp-ALA5-nnnn} alone, 
which can already give not only the precise LO KK graviton 
(KK Goldstone) amplitude, 
but also the {\it correct structure} of the NLO 
KK graviton (KK Goldstone) amplitude \eqref{eq:dM-dmT-DC}.
In the following, 
we will propose another improved double-copy method 
to further reproduce the exact longitudinal KK graviton 
(KK Goldstone) amplitudes at the NLO and beyond.\  
It only uses the amplitudes of pure longitudinal KK 
gauge bosons (KK Goldstone bosons) alone, 
hence it is practically simple and valuable.\
For this, we construct the following improved NLO numerators:
\beqs
\label{eq:dN'dNT'-XzzT}
\begin{align}
\hspace*{-3mm}
(\dNN_s'',\,\dNN_t'',\,\dNN_u'') &=
(\dNN_s^{\pp},\,\dNN_t^{\pp}\!-\!z,\,\dNN_u^\pp\!+\!z\,)\hs ,
\\[1mm]
\hspace*{-3mm}
(\dNNt_s'',\,\dNNt_t'',\,\dNNt_u'') &=
(\dNNt_s^{\pp},\,\dNNt_t^{\pp}\!-\!\zT,\,\dNNt_u^\pp \!+\!\zT\,)\hs ,
\end{align}
\eeqs
where 
$(z,\,\zT)$ are functions of $\,\theta\,$
and can be determined by matching our improved NLO KK amplitudes
of double-copy with the original NLO KK graviton (Goldstone) 
amplitudes of the GR5.\
Then, we solve $(z,\,\zT)$ as
%
\beqs
\label{eq:sol-z-zT} 
\begin{align}
\label{eq:sol-z}
z &\,=\,
\frac{\,M_n^2(1390 \!+\! 603\ctt\!+\! 66\ctf\!-11\cts)\,}
{12\hs (13\!-\!12\ctt \!-\! \ctf )}\,,
\\[0mm]
\label{eq:sol-zT}
\zT &\,=\,
\frac{\,M_n^2(4546 \!-\! 3585\ctt\!+\! 1086\ctf\!+\cts)\,}
{12\(13\!-\!12\ctt \!-\! \ctf \)}
\,.
\end{align}
\eeqs
%
Note that the modified kinematic numerators \eqref{eq:dN'dNT'-XzzT}
continue to hold the Jacobi identity.\
Because the corresponding NLO gauge (Goldstone) amplitudes 
$(\dT'',\,\da\tT'')$ are modified only by terms of NLO,
so we can still hold the general GAET identity 
$\hs\T'' \!=\tT''\!+\T_v''\hs$
by redefining the residual term as
$\,\T_v''\!=\T_v^{}\!-g^2
 (\CC_t/t\!-\hsm\CC_u/u)(z\!-\!\zT )\hs$.\ 
Using Eqs.\eqref{eq:dN'dNT'-XzzT}-\eqref{eq:sol-z-zT},
we can reproduce the exact NLO KK
gravitational scattering amplitudes 
[shown in 
 Eqs.\eqref{eq:dM-dMT-GR5xx}-\eqref{eq:Amp-E-2hLxx} 
 of the Supplemental Material\,\cite{supp}].\ 
This double-copy procedure can be further applied to  
higher orders beyond the NLO when needed.

\section{\hspace*{-2.5mm}GRET Residual Terms and Energy Cancellation}
\label{sec:6}
\vspace*{-2mm}


According to Table\,\ref{tab:1} and the generalized 
gauge transformation \eqref{eq:GGtransf}, 
we can explicitly deduce the equivalence between 
the KK gauge boson amplitude and the corresponding 
KK Goldstone boson amplitude:
\beq
\label{eq:GAET-LO}
\T_0^{}\,=\,\tT_0^{}\,,
\eeq 
\\[-5mm]
which belongs to the LO of $\mO(E^0M_n^0)$.
Using our double-copy method, we further derived the GRET
relation $\M_0^{}\!=\! \MT_0^{}$ at the $\hs\mO(E^2M_n^0)\hs$
as shown in Eq.\,\eqref{eq:ML0=M50}. 
Thus, the residual terms of the GAET
and the GRET \eqref{eq:GET} are given by the differences
between the KK longitudinal amplitude and KK Goldstone amplitude 
at the NLO$\hs$:
%
%
\beqs
\label{eq:Tv-Mdelta}
\begin{align}
\label{eq:Tv-count}
\hspace*{-3mm}
\T_v^{} &\equiv
\sum \!\T[\Afn , v_n] =
\dT \!-\! \da \tT
\,=\, \mO(\Mnn/E^2) \,,
\\[1mm]
\hspace*{-3mm}
\M_\Delta^{} &\equiv
\sum\!\M[\DEn,\phin] =
\dM \!-\! \dMT
= \mO(E^0\Mnn) \,.
\label{eq:Mdelta}
\end{align}
\eeqs
\\[-4mm]
The size of $\,\T_v^{}=\mO(\Mnn/E^2)\,$ can be easily understood
by using our generalized power counting rule\,\cite{supp}.\ 
But, making the direct power counting gives 
$\,\M_\Delta^{}\!=\hsm\mO(E^2)\,$  for its individual 
amplitudes, which has the same energy dependence
as the LO KK Goldstone amplitude \eqref{eq:AmpExHLO}.

\vspace*{1mm}

We can further determine the size 
of the residual term $\,\M_\Delta^{}$\, 
by the double-copy construction \eqref{AmpK-hLh5-nnnn} 
based upon the KK gauge (Goldstone) boson scattering amplitudes of
the YM5 theory alone (which are well understood\,\cite{5DYM2002}\cite{5DYM2002-2}\cite{KK-ET-He2004}\cite{5dSM}).\ From Eq.\eqref{AmpK-hLh5-nnnn} and Table\,\ref{tab:1}, 
we can estimate the residual term by power counting:
\\[-4mm]
\begin{align} 
\hspace*{-2mm} 
\M_\Delta^{}&=\,\mO(\da\M,\hs\da\MT)
=\mO\!\(\!
\frac{\NN_{\hsm j}^{0\hs\pp}\dNN_{\hsm j}^{\pp}}{s_{j}^{}},\,
\frac{\NNt_{\hsm j}^{0\hs\pp}\dNNt_{\hsm j}^{\pp}}{s_{j}^{}}\!\)
\nn\\[-1mm]
&=\, \mO(E^0\Mnn) \hs.
\end{align}
\\[-6mm]
Thus, we deduce the double-copy correspondence between the
residual term $\hs\T_v^{}\hs$ of the GAET and the residual term
$\M_\Delta^{}$ of the GRET:
\vspace*{-1mm}
\begin{equation}
\label{eq:RTerm}
\T_v^{}~\longrightarrow~\M_\Delta^{}(\rm{DC})
= \mO(E^0\Mnn)\hs .
\end{equation}
\\[-6mm]
Hence, our double-copy construction proves 
that the GRET residual term $\,\M_\Delta^{}$\, 
should have an energy cancellation
$\,\mO(E^2)\ito\mO(E^0)\,$ among its individual amplitudes
in Eq.\eqref{eq:GET-RT}.\
This proves that $\M_\Delta^{}$ is much smaller than the
leading KK Goldstone amplitude $\,\MT_0^{}=\mO(E^2M_n^0)$\,
under the high energy expansion.

\vspace*{1mm}

From the above double-copy construction,
we can establish a {\it new correspondence} 
from the GAET of the KK YM5 theory to the GRET of the
5d KK GR (GR5):
\begin{equation}
\label{eq:KKET-GET}
\rm{GAET\,(YM5)}
\ \Longrightarrow \
\rm{GRET\,(GR5)}\,.
\end{equation}
\\[-5mm]
We will give a systematically expanded analysis in the 
companison long paper\,\cite{Hang:2021fmp},  
which includes our elaborations of the current key points and 
our extension of KLT relations\,\cite{KLT}
(along with CHY\,\cite{CHY}) to the double-copy construction
of massive KK graviton amplitudes.

\vspace*{-2mm}
\section{\hspace*{-2.5mm}Conclusions}
\label{sec:7}
\vspace*{-2mm}

In this work, we newly formulated the geometric ``Higgs'' mechanism
for the mass generation of Kaluza-Klein (KK) gravitons
of the compactified 5d GR (GR5) theory 
at both the Lagrangian level and the scattering $S$-matrix level.\
Using a general $R_\xi^{}$ gauge-fixing of quantization, 
we proved that the KK graviton propagator
is {\it free from the longstanding problem} of the vDVZ discontinuity\,\cite{vDVZ}
in the conventional Fierz-Pauli massive gravity\,\cite{PF}\cite{Hinterbichler:2012}
and demonstrated that the KK gravity theory can {\it consistently
realize the mass-generation for spin-2 KK gravitons.}\

\vspace*{1mm}

We newly proposed and proved a
Gravitational Equivalence Theorem (GRET)
which connects the $N$-point scattering amplitudes of the longitudinal
KK gravitons to that of the gravitational KK Goldstone bosons.\
We computed the four-point scattering amplitudes of
KK Goldstone bosons in comparison with
the longitudinal KK graviton amplitudes,
and explicitly proved the equivalence between
the leading amplitudes of the longitudinal KK graviton
scattering and the corresponding KK Goldstone boson scattering
at $\mO(E^2M_n^0)$.\

\vspace*{1mm}

We developed a generalized power counting method
for massive KK gauge and gravity theories.\
Using the GRET and the new power counting rules, we established
{\it a general energy-cancellation mechanism} under which
the leading energy dependence of $N$-particle longitudinal KK graviton amplitudes
($\propto\! E^{2(N+1+L)}$)
must cancel down to a much lower energy power
($\propto\! E^{2(1+L)}$) by an energy factor of $E^{2N}$,
where $L$ denotes the loop number of the relevant Feynman diagram.\
For the case of longitudinal KK graviton scattering amplitudes with $N\!=\!4$ and $L\!=\!0$\,,
this proves the energy cancellations of $\,E^{10}\!\ito E^2\hs$.

\vspace*{1mm}

Extending the conventional massless double-copy
method\,\cite{BCJ:2008}\cite{BCJ:2019}
to the compactified massive KK YM and KK GR theories,\
we derived the Jacobi-respecting numerators and constructed
the scattering amplitudes of longitudinal KK gravitons
(KK Goldstone bosons) {under high energy expansion}.\
Using our extended massive double-copy approach, we 
constructed exact double-copy of the
KK graviton scattering amplitudes at both the 
leading order (LO) and the next-to-leading order (NLO).\ 
Applying this massive double-copy method,
we established {\it a new correspondence between the two
energy cancellations} in the four-point longitudinal KK
amplitudes: $E^4\!\ito E^0\,$ in the 5d KK YM gauge theory
and $\,E^{10}\!\!\to\! E^2\,$ in the 5d KK GR theory,
which is connected to the double-copy correspondence
between the GAET and GRET
as we derived in Eq.\eqref{eq:KKET-GET}.\ 
Furthermore, we analyzed the structure of the residual term
$\,\M_{\!\Delta}^{}$ in the GRET \eqref{eq:GET}
and further uncovered a new energy-cancellation mechanism
of $\,E^2\ito E^0\,$ for the residual term of the GRET.\

\vspace*{1mm} 

Finally, we stress that the geometric Higgs mechanism is a general
consequence of the KK compactification of extra spatial dimensions
and should be realized for other KK gravity theories 
with more than one extra dimensions or 
with nonflat extra dimensions.\
We note that our identity \eqref{eq:F-identityP}
results from the underlying gravitational diffeomorphism invariance
and thus should generally hold for any compactified 5d KK GR theory 
with proper gauge-fixing functions.\ 
Thus, we expect that the GRET should generally hold for
other 5d KK GR theories and take similar form as the present 
Eq.\eqref{eq:GET} \cite{GET-2}.\ For instance, we find that  
the geometric Higgs mechanism and the large energy-cancellations of the longitudinal KK graviton amplitudes are
also realized in the compactified  
warped 5d space of the Randall-Sundrum model\,\cite{RS}
and our GRET will work in the similar way.\ 
Following the current work,
it is encouraging to further study these interesting issues  
in our future work\,\cite{GET-2}.
In passing, we recently proposed\,\cite{Hang:2021oso} a brand-new
topological equivalence theorem (TET) to formulate
the topological mass-generation in the 3d topologically 
massive Yang-Mills theory (TMYM), with which we uncover 
the nontrivial energy cancellations in the $N$-point 
Chern-Simons scattering amplitudes of the massive physical  
gauge bosons, $E^4\ito E^{4-N}$.\ We then made an extended
double-copy construction of the four-point massive 
graviton scattering amplitude in the 3d 
topologically massive gravity (TMG)\,\cite{TMG}
and further proved\,\cite{Hang:2021oso} 
the striking energy cancellations of $\hs E^{12}\ito E^1\hs$ 
in such massive graviton scattering amplitude of the TMG
theory.\



\vspace*{3mm}

\noindent
{\bf Acknowledgements}
\\[1mm]
This research was supported in part 
by the National Natural Science Foundation
of China (under grants Nos.\,11835005 and 11675086), 
and by the National Key R\,\&\,D Program of China 
(under grant No.\,2017YFA 0402204).

\vspace*{4mm}

\noindent
{\bf Supplementary Materials}
\\[1mm]
In the following Supplementary Materials, 
we provide the relevant technical details for the analyses 
presented in the main text of this paper.
I.\,Kinematics of KK scattering;
II.\,Feynman rules for 5d KK GR theory;
III.\,Power counting and energy cancellations
for KK graviton amplitudes;
IV.\,KK graviton and Goldstone 
scattering amplitudes.



\clearpage
\newpage
\thispagestyle{empty}
\maketitle
\onecolumngrid

\begin{center}
\textbf{\large Gravitational Equivalence Theorem and Double-Copy
\\[0mm]
for Kaluza-Klein Graviton Scattering Amplitudes
\\[3mm]
--- Supplemental Material ---}
\\[5mm]
{\sc Yan-Feng Hang}\,$^1$ \,and\, {\sc Hong-Jian He}\,$^{1,2,3}$
\\[2mm]
$^1$\,Tsung-Dao Lee Institute $\&$ School of Physics and Astronomy,\\
Key Laboratory for Particle Astrophysics and Cosmology (MOE),\\
Shanghai Key Laboratory for Particle Physics and Cosmology,\\
Shanghai Jiao Tong University, Shanghai, China
\\[1mm]
$^2$\,Institute of Modern Physics $\&$ Physics Department,
Tsinghua University, Beijing, China
\\[1mm]
$^3$\,Center for High Energy Physics, Peking University, 
Beijing, China
\\[1mm]
({yfhang@sjtu.edu.cn, hjhe@sjtu.edu.cn})

\vspace{0.05in}
\end{center}
\onecolumngrid
\setcounter{equation}{0}
\setcounter{figure}{0}
\setcounter{table}{0}
\setcounter{section}{0}
\setcounter{page}{1}
\makeatletter
\renewcommand{\theequation}{S\arabic{equation}}
\renewcommand{\thefigure}{S\arabic{figure}}
\renewcommand{\thetable}{S\arabic{table}}

\vspace*{5mm}

This Supplemental Material provides in detail
the relevant formulas, Feynman rules, and the KK power counter method
for the analyses of the KK scattering amplitudes
in the compactified 5d Yang-Mills (YM5) theory and 
the compactified 5d General Relativity (GR5) theory.


\section{\hspace*{-2mm}Kinematics of KK Scattering}
\label{app:1}

We consider $\,2\ito 2\,$ KK scattering process,
with the four-momentum of each
external state obeying the on-shell condition
$\,p^2_j \!=\! -M_j^2$, ($j=1,2,3,4$).\ 
We number the external lines clockwise, with their momenta being
out-going. Thus, the energy-momentum conservation gives
$\,\sum_j p_j^{}\!=0\,$, and
the physical momenta of the two incident particles
equal $-p_1^{}$ and $-p_2^{}$, respectively.
For illustration, we take the elastic scattering
$\,X_n X_n\!\ito X_n X_n$ ($n\!\geqq\! 0$)
as an example,
where $X_n$ denotes any given KK state of level-$n$
and has $M_j^{}\!=\!M_n$.
For the KK theory, the external particle has
mass $M_n$ for a given KK-state of level-$n$.\
Thus, in the center-of-mass frame,
we define the momenta as follows:
\begin{equation}
\begin{alignedat}{3}
p_{1}^{\mu} &=  - E\(1, 0, 0, \be \) \,,
&& \hspace{10mm}
p_{2}^{\mu} = - E\(1, 0, 0, -\be \) \,,
\\[1mm]
p_{3}^{\mu} &=  E \( 1 , \be \st, 0, \be \ct  \) \,,
&&\hspace{10mm}
p_{4}^{\mu} =   E\( 1 , -\be \st, 0, -\be\ct \) \,,
\end{alignedat}
\end{equation}
where $(\st,\hs\ct)=(\sin\hsm\theta ,\, \cos\hsm\theta)$ and
$\hs\be\!=\!(1\!-\!\Mnn/\hsm E^2)^{1/2}\hs$.
With the above, we can define the following three Mandelstam variables:
\begin{equation}
\begin{aligned}
\label{eq:s-t-u}
s =-\( p_{1} \!+\hsm p_{2} \)^{2} \!= 4E^2 , ~~~~
t = -\( p_{1} \!+\hsm p_4 \)^{2} \!= -\fr{1}{2}s\hs\be^2
(1\!+\!\ct) \hs , ~~~~
u =-\( p_{1}^{} \!+\hsm p_3^{} \)^{2} \!=
-\fr{1}{2}s\hs\be^2 (1\!-\!\ct) \hs .
\end{aligned}
\end{equation}
Then, using the on-shell condition 
$E^2 \!=\hsm E^2\be^2\hsm +\hsm M_n^2\hs$, we define a new set 
of mass-independent Mandelstam variables as follows:
\\[-6mm]
\begin{align}
\label{eq:s0-t0-u0}
\sz = 4E^2\be^2  , \qquad
\tz = -\frac{\,\sz\,}{2}(1+\ct)  \hs , \qquad
\uz = -\frac{\,\sz\,}{2}(1-\ct)  \hs ,
\end{align}
where $\hs\sz\!=\! s\hsmx -\hsmx 4\Mnn$\,,
and thus $(\sz,\,\tz,\,\uz)=(s\be^2\hsm ,\,t,\,u)$.
Summing up the Mandelstam variables \eqref{eq:s-t-u}
and \eqref{eq:s0-t0-u0} gives the following relations:
\begin{equation}
\label{eq:stu-stu0}
s\hsm +t\hsm +u=4M_n^2\,, \qquad
\sz\hsm +\tz\hsm +\uz =0 \,.
\end{equation}

\vspace*{1mm}

As we mentioned in the text,
a massive KK graviton has 5 helicity states
($\lambda \!=\pm 2,\pm 1,0$\,). Their polarization tensors
take the following forms:
\begin{align}
\label{eq:hn-Pols}
\vep_{\pm 2}^{\mn} &=\ep_{\pm}^{\mu} \ep_{\pm}^{\nu}\,,\qquad
\vep_{\pm 1}^{\mn} =
\frac{1}{\sqrt{2\,}\,}\!\(\ep_{\pm}^{\mu}\ep_{L}^{\nu}
\!+\ep_{L}^{\mu} \ep_{\pm}^{\nu} \) \!,
\qquad
\vep_{L}^{\mn} \!=\frac{1}{\sqrt{6\,}\,}\!\( \ep_{+}^{\mu} \ep_{-}^{\nu}\!+\ep_{-}^{\mu} \ep_{+}^{\nu}\!
+2 \ep_{L}^{\mu} \ep_{L}^{\nu}\) \!,
\end{align}
where $(\ep_{\pm}^{\mu},\,\ep_{L}^{\mu})$
are the (transverse,\,longitudinal) polarization vectors
of a vector boson with the same 4-momentum $p^\mu$.
These polarization tensors obey the traceless and orthonormal
conditions. They are also orthogonal to the KK graviton's 4-momentum $p^\mu\hs$. Hence, the following conditions are realized:
\begin{equation}
\eta_{\mn}^{}\vep^{\mn}=0 \,, \qquad
\vep_\lambda^{\mn} \vep_{\lambda'\!,\,\mn}^*
= \delta_{\lambda\lambda'}^{} \,,\qquad
p_\mu^{} \vep^{\mn}=0 \,,
\end{equation}
where the helicity indices of each KK graviton are 
$\,\lambda,\lambda'=\pm 2,\hs\pm 1,\hs 0\,$.

\section{\hspace*{-2mm}Feynman Rules for 5d KK GR Theory}
\label{app:2}

In this section, we summarize the relevant
Feynman rules\,\cite{Hang:2021fmp}
including propagators and vertices
which are used for the amplitude calculations
in the text of this Letter.

\vspace*{1mm}

We first give the propagators in $R_\xi^{}$ gauge for KK graviton ($h^{\mn}_n$) and KK Goldstone bosons
$(\A^{\mu}_n,\,\phin)$ as follows:
\beqs
\label{eq:KKpropagator-Rxi}
\begin{align}
\hspace*{-2mm}
\D_{nm}^{\mn\ab}(p) \,=\,&
- \! \frac{\,\ii\delta_{nm}^{}\,}{2} \LB\!
\frac{\eta^{\mu \al}\eta^{\nu \be}\!+\!\eta^{\mu \be}\eta^{\nu\al}
\!-\!\eta^{\mu\nu}\eta^{\ab}}{p^{2}\!+\!M_{n}^{2}}
+\frac{1}{3}\!\[\!\frac{1}{\,p^2\!+\!M_n^2}\hsm -\hsm 
\frac{1}{\,p^{2}\!+\!(3\xi_n\!-\!2)M_n^{2}}
\!\] \!\!
\(\! \eta^{\mn}\!\hsm -\!\frac{\,2p^{\mu}_{}p^{\nu}_{}}{M_n^{2}} \!\)\!\!
\(\!\eta^{\ab}\!\hsm -\!\frac{\,2p^{\al}_{}p^{\be}_{}}{M_n^{2}} \!\)
\right. \nn\\[1mm]
& + \! \frac{1}{\,M_n^2\,}\!
\(\!\frac{1}{\,p^2\!+\!M_n^2} -
\frac{1}{\,p^{2}\!+\!\xi_n M_n^{2}\,}
\!\)\!
(\eta^{\mu\al}_{}p^{\nu}_{}p^{\be}_{}\!
+\! \eta^{\mu\be}_{}p^{\nu}_{}p^{\al}_{}\!
+\! \eta^{\nu\al}_{}p^{\mu}_{}p^{\be}_{}\!
+\! \eta^{\nu\be}_{}p^{\mu}_{}p^{\al}_{})
\nn\\[1mm]
& \left.
+\frac{\,4p^\mu_{}p^\nu_{}p^\al_{}p^\be_{}\,}{\xi_n^{}M_n^4}
\!\(\!\frac{1}{\,p^2\!+\!\xi_n^2M_n^2\,}
-\frac{1}{\,p^2\!+\!\xi_nM_n^2\,} \!\) \!\right\}\!,
\\[2mm]
\D^{\mn}_{nm}(p)  \,=\,&
\frac{-\ii\delta_{nm}^{}}{\,\,p^{2}\!+\!\xi_nM_{n}^{2}\,\,}\!
\left[\eta^{\mn} \!\!-\! \frac{\,p^{\mu} p^{\nu} (1\!-\!\xi_n)\,}
{\,p^{2}\!+\xi_n^2 M_{n}^{2}} \right]\!, \qquad
\D_{nm}^{}(p) =
\frac{-\ii\delta_{nm}^{}}{~p^2 \!+\! (3 \xi_n \!\!-\!2)M_n^2~} \,.
\end{align}
\eeqs
For the Feynman-'t\,Hooft gauge ($\xi_n^{} \!=\! 1$), the above
propagators reduce to the simple forms
[cf.\,\eqrefe{eq:D-xi=1} in the main text].

\vspace*{1mm}

Next, we make the following Fourier expansions
for the 5d graviton fields in terms of their zero modes
and KK states:
\beqs
\begin{align}
\hh^{\mn}(x^\rho, x^5) &\,=\,
\frac{1}{\sqrt{L\,}\,} \! \[  h^{\mn}_{0} (x^\rho)+
\sqrt{2} \sum_{n=1}^{\infty} h^{\mn}_{n}(x^\rho)
\cos\!\frac{n\pi x^5}{L} \!\] \!,
\label{eqHExp}
\\[1mm]
\hh^{\mu 5}(x^\rho, x^5) &\,= \,
\sqrt{\frac{2}{L}\,}
\sum_{n=1}^{\infty} h^{\mu 5}_{n}(x^\rho)
\sin\!\frac{n\pi x^5}{L}  \,,
\label{eqAExp}
\\[1mm]
\phih (x^\rho, x^5) &\,=\,
\frac{1}{\sqrt{L\,}\,} \[\phi_0 (x^\rho)+
\sqrt{2}\sum_{n=1}^{\infty} \phi_{n}(x^\rho)
\cos\!\frac{n \pi x^5}{L} \!\] \!.
\label{eqPhiExp}
\end{align}
\eeqs
With these, we list the relevant 4d effective Lagrangians
including both cubic and quartic interactions
which are used for our analyses:
\\[-4mm]
\beqs
\label{eq:LagKK}
\begin{align}
\La_{1} [ h\phi^2 ] =&\, \frac{\ka}{\sqrt{2\,}\,}\!\!  \sum_{n,m,\ell = 1}^{\infty}  \!\Bigl\{
a_1^{}\big[ \sqrt{2} \hs ( h_0^{\mn}  \pd_\mu \phi_0  \pd_\nu \phi_0  \!+\! h_0^{\mn} \pd_\mu \phi_m \pd_\nu \phi_{\ell}  \delta_{m\ell} \!+\!  h_n^{\mn} \pd_\mu \phi_m  \pd_\nu \phi_0  \delta_{n m} \!+\!  h_n^{\mn}\pd_\mu \phi_0 \pd_\nu \phi_{\ell}  \delta_{n\ell} ) 
\nn\\ 
&+ h_n^{\mn} \pd_\mu \phi_m \pd_\nu\phi_{\ell} \hs \Delta_3(n,m,\ell) \big]  
+ a_2^{} \big[ \sqrt{2} \hs (  h_0^{\mn}  \phi_0  \pd_\mu\pd_\nu \phi_0 \!+\! h_0^{\mn} \phi_m  \pd_\mu\pd_\nu \phi_{\ell}  \delta_{m\ell} \!+\!  h_n^{\mn}  \phi_m  \pd_\mu\pd_\nu \phi_0  \delta_{nm} 
\nn\\[0mm]
&+ h_n^{\mn} \phi_0  \pd_\mu\pd_\nu \phi_{\ell}  \delta_{n\ell} ) 
\!+\!  h_n^{\mn} \phi_m \pd_\mu\pd_\nu \phi_{\ell} \Delta_3(n,m,\ell) \big]
\!+ a_3^{} \big[ \sqrt{2} ( h_0 \pd_\mu \phi_0 \pd^\mu \phi_0  \!+\! h_0 \pd_\mu \phi_m \pd^\mu \phi_{\ell} \delta_{m\ell}
\nn\\[0mm]
&+ h_n \pd_\mu \phi_m \pd^\mu \phi_0 \delta_{n m} 
\!+\!  h_n \pd_\mu \phi_0 \pd^\mu \phi_{\ell}  \delta_{n\ell} )  \!+\!  h_n \pd_\mu \phi_m \pd^\mu\phi_{\ell}\hs \Delta_3(n,m,\ell) \big]
\!+ a_4^{}  \big[\sqrt{2}\hs ( h_0  \phi_0  \pd^2_\mu \phi_0 
\nn\\[0mm]
&+ h_0  \phi_m \pd^2_\mu \phi_{\ell} \delta_{m\ell} 
+ h_n \phi_m  \pd^2_\mu\phi_0  \delta_{nm} \!+\! h_{n}   \phi_0  \pd^2_\mu \phi_{\ell}  \delta_{n\ell})
\!+\! h_n \phi_m  \pd^2_\mu \phi_{\ell}\hs\Delta_3(n,m,\ell) \big]
\nn\\
&+ a_5^{} \hs M_m^{} M_\ell^{} 
\big[ 
\sqrt{2}\hs h_0 \phi_m \phi_{\ell} \delta_{m\ell}
\!+\! h_n\phi_m\phi_{\ell}\hs \widetilde{\Delta}_3(n,m,\ell)\big]
\!- a_6^{} M_\ell^2 \big[ \sqrt{2} \hs ( h_0 \phi_m  \phi_{\ell}  \delta_{m\ell} \!+\! h_n \phi_0 \phi_{\ell} \delta_{n\ell} ) 
\nn\\
&+ h_n\phi_m\phi_{\ell}\hs \Delta_3 (n,m,\ell) \big]   \Bigr\}\,,
\\[2mm]
\La_1[\A \hs \phi^2 ]
=&\ - \frac{\ka}{\sqrt{2}}  \sum_{n,m,\ell=1}^{\infty} \! \Bigl\{  
b_1^{} \hs M_\ell \big[  
\sqrt{2} \hs \A^{\mu}_n  \pd_\mu \phi_0  \phi_\ell \delta_{n\ell} 
\hsm +\hsmx  \A^{\mu}_n  \pd_\mu \phi_m  \phi_\ell \hs \widetilde{\Delta}^{\pp}_3(n,m,\ell) \big]
+ b_2^{} \hs M_\ell \big[ \sqrt{2}\hs \A^{\mu}_n \phi_0  \pd_\mu \phi_\ell  \delta_{n\ell}
\nn\\[-1mm]
&
+ \A^{\mu}_n  \phi_m  \pd_\mu \phi_\ell \hs \widetilde{\Delta}^{\pp}_3 (n,m,\ell) \big] \Bigr\}  \,,
\\[2mm]
\La_{1} [ \phi^3 ] =&\,
\frac{\ka}{\sqrt{2}}  \sum_{n,m,\ell=1}^{\infty}\!
\Big\{  c_1 \big[ \sqrt{2} \hs ( \phi_0 (\pd_\mu\phi_0)^2  \!+\! \phi_0 \pd_\mu \phi_m \pd^\mu \phi_\ell \, \delta_{m\ell} \!+\! \phi_n \pd_\mu \phi_0 \pd^\mu \phi_m \delta_{n m}   
\!+\!  \phi_n  \pd_\mu \phi_0 \pd^\mu \phi_\ell \, \delta_{n\ell} ) 
\nn\\[-1mm]
&+ \phi_n \pd_\mu \phi_m \pd^\mu \phi_\ell \hs \Delta_3(n,m,\ell) \Big] \!+ c_2 M_m M_\ell \big[ \sqrt{2} \, \phi_0\phi_m\phi_\ell  \delta_{m\ell} \!+\!  \phi_n\phi_m\phi_\ell  \widetilde{ \Delta}_3(n,m,\ell)  \big] \Big\}  \,,   
\\[2mm]
\La_2 [ \phi^4 ] =&\,
\frac{\ka^2}{2}\! \sum_{n,m,\ell,k=1}^{\infty} \!\Bigl\{ d_1^{}\! \left\{ 2(\phi_{0}\pd_{\mu} \phi_{0})^{2} \!+\! 2\big[ \! \(\pd_\mu \phi_{0}\)^{2} \phi_{n} \phi_{m} \delta_{n m}\!+\!\phi_{0} \pd_{\mu} \phi_{0} \phi_{n} \pd^{\mu} \phi_{\ell} \delta_{n\ell}  \!+\! \phi_{0} \pd_{\mu} \phi_{0} \phi_{n} \pd^{\mu} \phi_{k} \delta_{nk}
\right.  \nn\\[-0.5mm]
&\!+  \phi_{0} \pd_{\mu} \phi_{0} \phi_{m} \pd^{\mu} \phi_{k} \delta_{mk}  \!+\!  \phi_{0} \pd_{\mu} \phi_{0} \phi_{m} \pd^{\mu} \phi_\ell \delta_{m\ell} \!+\! (\phi_{0} )^{2} \pd_{\mu} \phi_\ell \pd^{\mu} \phi_{k}  \delta_{\ell k}  \big]
\!+\! \sqrt{2}  \[ \pd_{\mu} \phi_{0} \phi_{n} \phi_{m} \pd^{\mu} \phi_\ell  \Delta_{3}(n, m, \ell) \right.
\nn\\[1mm]
&\left. \!+\,  \pd_{\mu} \phi_{0} \phi_{n} \phi_{m} \pd^{\mu} \phi_{k} \hs \Delta_{3}(n, m, k)\!+\!\pd_{\mu} \phi_{0} \phi_{n} \phi_\ell \pd^{\mu} \phi_{k} \Delta_{3}(n, \ell, k)   \!+\! \phi_{0} \phi_{m} \pd_{\mu} \phi_\ell \pd^{\mu} \phi_{k}  \Delta_{3}(m, \ell, k) \]  \nn\\[1mm]
&\left. \!+\,  \phi_{n} \phi_{m} \pd_{\mu} \phi_\ell \pd^{\mu} \phi_{k} \hs\Delta_{4}(n, m, \ell, k ) \right\}
\!+ d_2^{}  M_\ell M_{k} \big[ 2 (\phi_{0})^2 \phi_\ell \phi_{k} \delta_{\ell k}\!+\! \sqrt{2}  \phi_{0} \phi_{m} \phi_\ell \phi_{k}  \widetilde{\Delta}_{3}(m, \ell, k)   
\nn \\ 
& \!+ \sqrt{2} \hs \phi_{0} \phi_{n} \phi_\ell \phi_{k} \hs \widetilde{\Delta}_{3}(n, \ell, k)
\!+\! \phi_{n} \phi_{m} \phi_\ell \phi_{k} \hs \widetilde{\Delta}_4(n, m, \ell, k) \big] \Bigr\} \,,
\end{align}
\eeqs
where the delta functions ($\Delta_j^{},\widetilde{\Delta}_j^{}$) are defined as follows:
\begin{equation}
\begin{aligned}
\Delta_3(n,m,\ell)  \,=\,&\ \delta(n\!+\!m\!-\!\ell)\!+\!\delta(n\!-\!m\!-\!\ell)\!+\!\delta(n\!-\!m\!+\!\ell) \,,
\\[1mm]
\widetilde{\Delta}_3(n,m,\ell) \,=\,&\ \delta(n\!+\!m\!-\!\ell)\!-\!\delta(n\!-\!m\!-\!\ell)\!+\!\delta(n\!-\!m\!+\!\ell)  \,,
\\[1mm]
\widetilde{ \Delta}_3^{\prime}(n,m,\ell)  \,=\,&\  \delta(n\!+\!m\!-\!\ell)\!-\!\delta(n\!-\!m\!+\!\ell)\!+\!\delta(n\!-\!m\!-\!\ell)  \,,
\\[1mm]
\Delta_4(n,m,\ell,k) \,=\,& \ \delta(n\!+\!m\!+\!\ell\!-\!k)\!+\!\delta(n\!+\!m\!-\!\ell\!-\!k)\!+\!\delta(n\!-\!m\!+\!\ell\!-\!k) \!+\! \delta(n\!-\!m\!-\!\ell\!-\!k) \\
&\ \!+\!\delta(n\!-\!m\!-\!\ell\!+\!k)  \!+\!\delta(n\!+\!m\!-\!\ell\!+\!k) \!+\!\delta(n\!-\!m\!+\!\ell\!+\!k) \,,
\\[1mm]
\widetilde{\Delta}_4(n,m,\ell,k) \,=\,& \ \delta(n\!+\!m\!+\!\ell\!-\!k)\!-\!\delta(n\!+\!m\!-\!\ell\!-\!k)\!+\!\delta(n\!-\!m\!+\!\ell\!-\!k)\!-\! \delta(n\!-\!m\!-\! \ell\!-\!k)  \\
&\ \!+\!\delta(n\!-\!m\!-\!\ell\!+\!k)   \!-\! \delta(n\!+\!m\!-\!\ell\!+\!k) \!+\!\delta(n\!-\!m\!+\!\ell \!+\!k)   \,.
\end{aligned}
\end{equation}
%
Then, we derive the Feynman rules based on the interaction
Lagrangians in the above \eqrefe{eq:LagKK}.\
We present the relevant 3-point and 4-point vertices as follows:
\vspace*{1mm}
\beqs
\begin{align}
\raisebox{0.em}{\hbox{\begin{minipage}{3.8cm}
\hspace*{-16mm}
\includegraphics[height=3.cm]{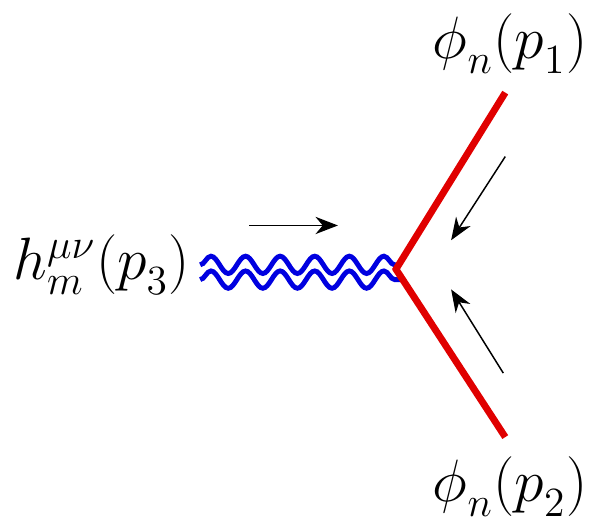}
\end{minipage}}}
\hspace*{-16mm}
\hspace{3mm}&= \frac{-\ii\ka}{\sqrt{1\!+\!\delta_{2n,m}\,}\,} \!\!
\[ \hspace{-.3em}
\begin{array}{cl}
& \hspace{-1.5mm}
a_1^{} \! \( p_1^{\mu} p_2^{\nu} \!+\! p_1^{\nu} p_2^{\mu} \)
\\[1.5mm]
+&\hspace{-1.5mm}
a_2^{} \!\( p_1^{\mu} p_1^{\nu} \!+\! p_2^{\mu} p_2^{\nu} \)
\\[1.5mm]
+&\hspace{-1.5mm}
2 \hs a_3^{} \eta^{\mn} \!\(p_1^{}  \!\cdot p_2^{}\)
\\[1.5mm]
-&\hspace{-1.5mm}
2\hs \tilde{a}_4^{} \eta^{\mn} M_n^2
\\[.5mm]
\end{array} \! \]\!,
\label{eq:VertexHPP}
\\[1.3mm]
\raisebox{0.em}{\hbox{\begin{minipage}{4.8cm}
\centering
\includegraphics[height=3.cm]{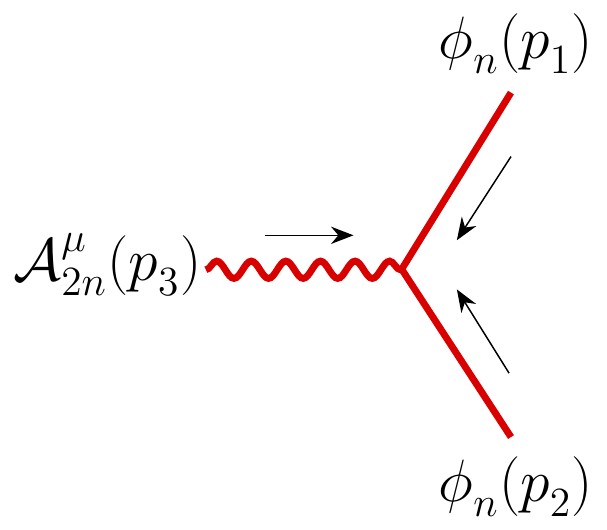}
\end{minipage}}}
\hspace{-0.4mm}
&
= \ - \frac{\,\ka (b_1^{}\!+b_2^{})\Mn\,}{\sqrt{2\,}}
(p_1^{\mu} \!+ p_2^{\mu})  \,,
\\[1.3mm]
\raisebox{0.em}{\hbox{\begin{minipage}{4.5cm}
\centering
\includegraphics[height=3.cm]{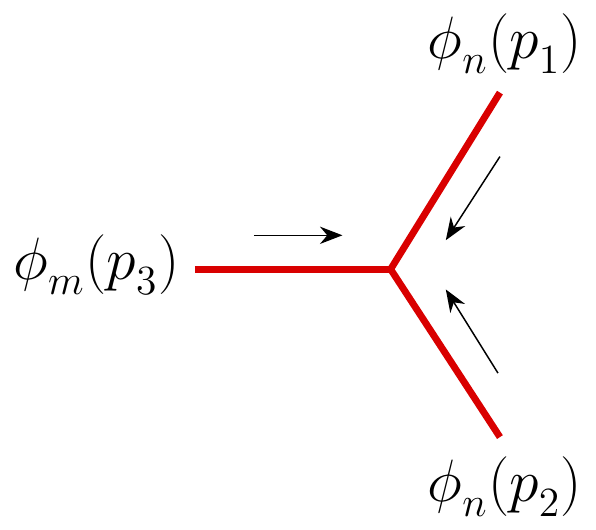}
\end{minipage}}}
\hspace{.5mm}
&=\LB
\begin{aligned}
m &= 0\!: \ \ii\,2\ka \!
\[\!c_1^{} (p_1^2 + p_2^2 + p_1^{} \!\cdot p_2^{}) \!+\hsmx c_2^{} M^2_n \]
\\[.5mm]
&\hspace{0.9cm} \xrightarrow{\rm{on}\text{-}\rm{shell}}
\ii\,2\ka \!
\[\!c_1^{}(p_1^{}\!\cdot p_2) \!-\hsmx (2c_1^{}\!-\hsmx c_2^{})M^2_n \] \!,
\\[1.mm]
m &= 2n\!: - \ii\sqrt{2}\,\ka\!
\[\!c_1^{} (p_1^{} \!\cdot p_2^{})\hsmx +c_2^{} M^2_n \] \!,
\end{aligned}
\right.
\\[1.3mm]
\raisebox{0.em}{\hbox{\begin{minipage}{4.5cm}
\centering
\includegraphics[height=3cm]{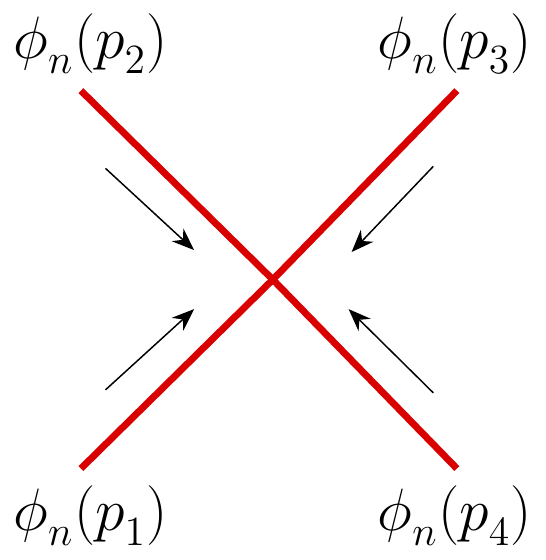}
\end{minipage}}}
&
\begin{aligned}
&= \ii\,6\ka^2 \! \[\! d_1^{} (p_1^2 + p_2^2 + p_1^{}
\!\cdot p_2^{} - p_3^{} \!\cdot p_4^{}) +2 d_2^{} M_n^2 \]
\\[1mm]
&\hspace*{4mm}
\xrightarrow{\rm{on}\text{-}\rm{shell}}
-\ii\,12\ka^2(d_1^{}\!-d_2^{})M_n^2  \,.
\end{aligned}
\end{align}
\eeqs
where
$\,\tilde{a}_4 = a_4^{} + (-1)^{\delta_{2n,m}} a_5^{}
-a_6^{}$\, with $m\!=0,\,2n$\, in \eqrefe{eq:VertexHPP}.

\section{\hspace*{-2mm}Power Counting and Energy Cancellations
for KK Graviton Scattering Amplitudes}
\label{app:3}

We consider a $S$-matrix element $\,\mathbb{S}\,$ having
$\,\EE\,$ external states and $L$ loops ($L\!\geqq\! 0$).\ 
Extending the original power counting rule for the 
ungauged nonlinear $\sigma$-model of the 
low energy QCD by Steven Weinberg\,\cite{weinberg}\cite{steve-foot},
we develop generalized power counting approach\,\cite{Hang:2021fmp} 
for the KK gravity theory.\ The mass dimension of a given 
scattering amplitude $\,\Sb\,$ in 4d is counted as
\\[-5mm]
\begin{equation}
\label{eq:DS}
D_{\mathbb{S}}^{} \,=\,4 - \mathcal{E}\,,
\end{equation}
where the number of external states
$\,\EE \!=\EE_B^{}+\EE_F^{}\,$
with $\,\EE_B^{}\,(\EE_F^{})\,$ representing the total number of
external bosonic (fermionic) states. In addition, we only consider the SM fermions whose masses are much smaller than the
scattering energy.
We denote the number of vertices of type-$j$ as $\VV_j^{}$\,.
Each vertex of type-$j$ contains $\,d_j^{}\,$ derivatives,
$\,b_j^{}\,$ bosonic lines and $\,f_j^{}\,$ fermionic lines.
Then, the energy dependence of coupling constant in $\,\mathbb{S}\,$ is given by
\begin{equation}
\label{eq:DC}
D_C^{} \,=\, \sum_j \VV_j^{}\!
\(4-d_j^{}\!-b_j^{}\!-\fr{3}{2}f_j^{}\) \!.
\end{equation}
For each Feynman diagram in the amplitude
\,$\mathbb{S}$\,,\,
we denote the number of the internal lines as
$\,I=I_B^{}+I_F^{}\,$ with
$\,I_B^{}$ ($\,I_F^{}\,$) being the number of the internal
bosonic (fermionic) lines. Thus, we have the following general
relations:
\begin{equation}
\label{eq:L-V-I}
L = 1+I-\VV\,, \hspace*{5mm}
\VV=\sum_j \VV_j^{} \,, \hspace*{5mm}
\sum_j \VV_j^{}b_j^{} = 2I_B^{}+\EE_B^{}\,, \hspace*{5mm}
\sum_j \VV_j^{}f_j^{} = 2I_F^{}+\EE_F^{}\,,
\end{equation}
where $\,\VV\,$
is the total number of vertices in a given Feynman diagram.\
The $\,\mathbb{S}\,$ may include $\,\EE_{h_L}^{}\!$
external longitudinal KK graviton states.
Thus, taking Eqs.\eqref{eq:DS}-\eqref{eq:L-V-I}, we deduce
the leading energy-power dependence as follows:
\begin{equation}
\label{eq:D_E}
D_E^{} \,=\, D_{\mathbb{S}}-D_C \,=\,
2\hs \EE_{h_L}^{} \!\!+
(2L\!+2)+\sum_j  \VV_j^{}\!
\(d_j^{}\!-2+\!\fr{1}{2}f_j^{}\) \!.
\end{equation}

For the pure longitudinal KK graviton scattering amplitude
with $\,N\,$ external states,
we have $\,\EE_{h_L}^{}\!\!\!=\!N\,$ and $\,f_j^{}\!=\!0\,$.\
Each pure KK graviton vertex always contains two partial derivatives
and thus $\,d_j^{}\!=2\,$.
For the loop level ($L\geqq 1$),
the amplitude may contain gravitational ghost loop which
involves graviton-ghost-antighost vertex, but the number of partial
derivatives $\,d_j^{}$ should be no more than two.\
While for the gravitational
KK Goldstone boson scattering
amplitude, its leading energy dependence is given by the diagrams
containing the cubic vertices of type $h^{\mn}_n$-$\phi_m^{}$-$\phi_\ell^{}$
and the pure graviton self-interaction vertices, where
each of these vertices includes two derivatives
($d_j^{}\!=2$).
Hence, we can derive the power counting
formula \eqref{eq:D_E} as:
%
\begin{equation}
\label{eq:DE-hL-phin}
D_E^{}[Nh_L^n] \,=\, 2(N\!+\!1)\!+2L\,, \qquad
D_E^{}[N\phin] \,=\, 2+2L\,,
\end{equation}
where the notation $\,[Nh_L^n]\,$ and $\,[N\phin]\,$ 
denote the $N$ external longitudinal KK
graviton states and $N$ external KK Goldstone states, respectively.

\vspace*{1mm}

Comparing the energy power counting formulas for KK graviton and KK Goldstone in \eqrefe{eq:DE-hL-phin}, we note that their difference arises from
the leading energy-dependence of the polarization tensors
$\,\vep_L^{\mn}\!\!\sim\!k^\mu k^\nu/\Mnn\,$ for the $N$ external
longitudinal KK gravitons in the high energy scattering:
\begin{equation}
\label{eq:DEL-DEphi}
D_E^{}[Nh_L^n] - D_E^{}[N\phin] \,=\, 2N \hs.
\end{equation}
Finally, we examine the leading energy dependence of the individual
amplitudes in the residual term $\,\MD\,$ of the GRET [cf.\,\eqrefe{eq:GET} in main text].\
A typical leading amplitude can be
$\M [\vt_{n_1}^{},\cdots\!, \vt_{n_N}^{}]\,$,\,
in which all the external states are KK gravitons
contracted with
$\,\vt^{\mn}\!\hsm =\hsm\vep_L^{\mn}\!\!-\vep_S^{\mn}\!=\mO(E^0)\,$,
such as $\,\vt_n^{}\!=\!\vt_{\mn}h_n^{\mn}\hs$.
Hence, the leading energy dependence of
this amplitude yields:
\\[-5.5mm]
\begin{equation}
\label{eq:DE-vn}
D_E^{}[N\vt_n^{}] \,=\, 2+2L\,,
\end{equation}
which gives the same energy power dependence as 
$\,D_E^{}[N\phin]\hs$.

\section{\hspace*{-2mm}KK Graviton and Goldstone 
Scattering Amplitudes}
\label{app:4}

In this section, we first present the four-point 
scattering amplitudes of KK gravitons (Goldstone bosons) 
at the LO and NLO of the high energy expansion,
which are obtained from computing the Feynman diagrams.\
Then, we present the four-point scattering amplitudes
of the KK gauge bosons (Goldstone bosons) at the LO and
NLO under two kinds of high energy expansions.
From these we provide the detailed formulas for our
improved massive double-copy construction of the KK graviton
(Goldstone) amplitudes which are used in the main text. 

\subsection{\hspace*{-2mm}KK Graviton and Goldstone Amplitudes
from Feynman Diagrams}
\label{app:4.1}

In this subsection, we summarize the full elastic amplitudes of
the four longitudinal KK graviton scattering\,\cite{Chivukula:2019L}
and of the four gravitational
KK Goldstone boson scattering\,\cite{Hang:2021fmp}.
For the purpose of our double-copy analysis, we express these
amplitudes in terms of the dimensionless variable $\,\bs$\,:
\beqs
\label{eq:AmpFull-4hL-4Phi}
\begin{align}
\M[4h_L^n] &= 
-\frac{~\ka^2 M_n^2(X_0 +\hsm X_2\ctt +\hsm X_4\ctf +\hsm X_6\cts )
\hsm\csc^2\!\theta~}
{512\hs\bs\hs (\bs-4)
[\hs\bs^2 -\hsm (\bs-4)^2 \ctt + 24\hs\bs +\hsm 16\hs ]}  \,,
\label{eq:FullAmp4hL}
\\[1.mm]
\MT[4\phin] &=
-\frac{~\ka^2 M_n^2(\tX_0+\tX_2\ctt +\tX_4\ctf +\tX_6\cts )\csc^2\!\theta~}
{512\hs\bs\hs (\bs-4)
[\hs\bs^2 -(\bs-4)^2 \ctt\hsm +\hsm 24 \bs +\hsm 16\hs]}  \,,
\label{eq:FullAmp4Phi}
\end{align}
\eeqs
where 
$\,\bs =s/\Mnn\,$ and $\,c_{n\theta}^{}\hsm =\cos(n\theta)\hs$. 
In the above, the coefficients $(X_j^{},\,\tX_j^{})$ 
are defined as follows:
\begin{equation}
\begin{aligned}
X_0 &= -2 (255 \bs^5 + 2824 \bs^4 - 19936 \bs^3 + 39936 \bs^2 - 256 \bs + 14336)\hs ,
\\[1mm]
X_2 &= 429 \bs^5-10152 \bs^4+30816 \bs^3-27136 \bs^2-49920 \bs+34816\,,
\\[1mm]
X_4 &= 2 (39 \bs^5 - 312 \bs^4 - 2784 \bs^3 - 11264 \bs^2 + 26368 \bs - 2048) \hs ,
\\[1mm]
X_6 &= 3 \bs^5+40 \bs^4+416 \bs^3-1536 \bs^2-3328 \bs-2048 \,,
\\[1.5mm]
\widetilde{X}_0 &=-2 (255 \bs^5+8248 \bs^4 \, -4144 \bs^3+79104 \bs^2+642560 \bs+69632)\hs,
\\[1mm]
\widetilde{X}_2 &=429 \bs^5+4152 \bs^4\, +21216 \bs^3-150016 \bs^2+1142016 \bs+182272 \,,
\\[1mm]
\widetilde{X}_4 &= 2(39 \bs^5-1992 \bs^4+17808 \bs^3-58112 \bs^2+70144 \bs-20480) \,,
\\[1mm]
\tX_6 &= 3 \bs^5-56 \bs^4+416 \bs^3-1536 \bs^2+2816 \bs-2048  \,.
\end{aligned}
\end{equation}
Then, we expand the KK graviton and KK Goldstone scattering amplitudes \eqref{eq:FullAmp4hL}-\eqref{eq:FullAmp4Phi}
under the high energy expansion of $\,1/s\,$:
\beqs
\begin{align}
\M[4\hLn] &\,=\,
\M_0[4h_L^n] + \dM[4h_L^n]  \,,
\\[1mm]
\MT[4\phin] &\,=\,
\MT_0[4\phin] + \dMT[4\phin] \,,
\end{align}
\eeqs
where the LO and NLO KK amplitudes take the following forms,  
\beqs
\begin{align}
\label{eq:dM-dMT-GR5xx} 
\M_0^{}[4h_L^n] &\,=\, \MT_0^{}[4\phin] = 
\frac{~3\ka^2\,}{~128~}\hs s\hs  
( 7\hsm +\ctt )^2\hsm \csc^2\!\theta  \,,
\\[1mm]
\label{eq:Amp-E-2hLxx} 
\da\M [4h_L^n] &\,=\, -\frac{~\ka^2 M_n^2~}{256}
(1810 \hsm +\hsm 93\hs\ctt \hsm +\hsm 126\hs\ctf \hsm +\hsm 
19\hs\cts)\hsm \csc^4 \hsmx\theta  \,,
\\[2mm]
\label{eq:Amp-E-2phixx}
\da\MT [4\phin ] &\,=\, -\frac{~\ka^2 M_n^2~}{256}   
(-\hs 902 \hsm+\hsm 3669 \ctt \hsm-\hsm 714 \ctf 
\hsm -\hsm 5\hs\cts ) \hsm\csc^4 \hsmx\theta\,.
\end{align}
\eeqs
If we make instead the high energy expansion in terms of 
$\hs 1/\sz\hs$,
we derive the following LO and NLO KK amplitudes:
\\[-4mm]
\beqs
\label{eq:dM-dMT-GR5x}
\begin{align}
\label{eq:M0=MT0}
\M_0^{\,\pp}[4h_L^n] &\,=\,	\MT_0^{\,\pp}[4\phin] 
\hs =\hs 
\frac{3 \ka^2}{~128~}\hs\sz\hs    
( 7+\ctt )^2\hsm \csc^2\!\theta  \,,
\\[1mm]
\delta\M^{\,\pp}[4h_L^n]  &\,=\,
-\frac{\,\ka^2 M_n^2~}{128}(650+261\ctt\!+102\ctf\!+11\cts )
\csc^4\!\theta   \,,
\label{eq:Amp-E-2hL}
\\[1mm]
\delta\MT^{\,\pp}[4\phin]  &\,=\,
-\frac{\,\ka^2 M_n^2~}{128}
(-706+2049 \ctt - 318 \ctf -\cts )
\csc^4\!\theta \,,
\label{eq:Amp-E-2phi}
\end{align}
\eeqs
where $\,\sz =s\hsm -\hsm 4\Mnn\hs$. 
We see that the $\hs 1/\sz\hs$ expansion has shifted 
a hidden $\mO(\Mnn)$ subleading term (contained in  
$\hs s= \sz + 4\Mnn\hs)\hs$
from the LO amplitudes \eqref{eq:dM-dMT-GR5xx} 
into the NLO amplitudes \eqref{eq:Amp-E-2hL}-\eqref{eq:Amp-E-2phi}.\
But this rearrangement in Eqs.\eqref{eq:M0=MT0}-\eqref{eq:Amp-E-2phi}
does not affect the difference between the two NLO amplitudes.\
Thus, we can deduce the contribution of the residual terms
by computing the amplitude-difference from either
Eqs.\eqref{eq:Amp-E-2hLxx}-\eqref{eq:Amp-E-2phixx} or  
Eqs.\eqref{eq:Amp-E-2hL}-\eqref{eq:Amp-E-2phi} as follows:
\begin{equation}
\label{eq:RTerm-NLO-GR}
\M^{}_{\Delta} =\, \delta \M[4h_L^n] -  \delta\MT[4\phin]
\,=\, 
-\frac{\,3\hs\ka^2\hsm M_n^2\,}{2}\!\(\! \frac{\,39\,}{2}+\ctt\!\) 
\!.
\end{equation}
This provides \eqrefe{eq:Diff2-AmpGR5-nnnn} 
in the main text.

\subsection{\hspace*{-2mm}KK Graviton and Goldstone Amplitudes
from Extended Double-Copy}
\label{app:4.2}

We expand the scattering amplitudes under 
the high energy expansion in terms of $\hs\Mnn/s\hs$.\ 
Thus, we can express 4-point elastic KK gauge boson
(Goldstone) amplitudes as follows:
\beqs
\label{xAmp0-ALA5-nnnn}
\begin{align}
\label{xAmp0-AL-nnnn}
\T[4A^{n}_{L}]
&\,=\, g^2\! \(\! \frac{\,\CC_s \NN_s\,}{s}
+ \frac{\,\CC_t \NN_t\,}{t}
+ \frac{\,\CC_u \NN_u\,}{u}  \!\)  \!,
\\[1mm]
\label{xAmp0-A5-nnnn}
\tT[4A^{n}_5]
&\,=\, g^2\! \(\! \frac{\,\CC_s \NNt_s\,}{s}
+ \frac{\,\CC_t \NNt_t\,}{t}
+ \frac{\,\CC_u \NNt_u\,}{u}  \!\)  \!,
\end{align}
\eeqs
which are invariant under the following generalized 
gauge-transformations:
\begin{equation}
\label{eq:xGGtransf}
\NN_{\hsm j}^{\pp} = \NN_{\hsm j}^{} + s_{\hsm j}^{}\hs\Delta \,,
\qquad
\NNt_{\hsm j}^{\pp} = \NNt_{\hsm j}^{} + s_{\hsm j}^{}\hs
\widetilde{\Delta} \,.
\end{equation}
The above Eqs.\eqref{xAmp0-ALA5-nnnn}-\eqref{eq:xGGtransf}
are given in 
Eqs.\eqref{Amp-ALA5-nnnn}\eqref{eq:GGtransf}
of the main text.\ 
This allows us to find proper solutions of 
$\{\Delta\,,\widetilde{\Delta}\}$ which ensure the
gauge-transformed NLO numerators 
$(\da\NN_j^{\pp},\,\da\NNt_j^{\pp})$
to obey the kinematic Jacobi identity,
as we demonstrated in 
Eqs.\eqref{eq:sol-Delta-tDelta}-\eqref{eq:sol-Delta01}
of the main text 
(cf.\ Sec.\ref{sec:5}).
Thus, from these we can derive the
gauge-transformed NLO numerators for the elastic KK gauge boson
amplitude:
\beqs
\label{eq:Nj'-NLO}
\begin{align}
\da\NN^{\pp}_s &= 
-\fr{1}{4}\Mnn\hs (246\hs\ct \!+\! 7\hs\cttt \!+\! 3\hs\ctfif)
\hsm\csc^4\!\theta \,, 
\\[1.5mm]
\da\NN^{\pp}_t &= 
\frac{~\Mnn\hs (131 \!-\! 8 \ct \!-\! 4 \ctt \!+\! 8\cttt \!+\! \ctf)~}
{8\hs (1\!-\!\ct)^2} \,,
\\ 
\da\NN^{\pp}_u &= 
-\frac{~\Mnn\hs (131 \!+\! 8 \ct \!-\! 4 \ctt \!-\! 8\cttt \!+\! \ctf)~}
{8\hs (1\!+\!\ct)^2} \,,
\end{align}
\eeqs
and the gauge-transformed NLO numerators for the
corresponding KK Goldstone boson amplitude:
\beqs
\label{eq:Ntj'-NLO}
\begin{align}
\da\NNt^{\pp}_s &= 
-\fr{1}{4}\Mnn\hs (238\hs\ct \!+\! 19\hs\cttt \!-\!  \ctfif)
\hsm\csc^4\!\theta \,, 
\\[1.5mm]
\da\NNt^{\pp}_t &= 
\frac{~\Mnn\hs (99 \hsm +\hsm 8\hs\ct \!+\! 28 \ctt 
\!-\hsm 8\hs\cttt \!+\hsm \ctf)}
{8\hs (1\!-\!\ct)^2}  \,,
\\ 
\da\NNt^{\pp}_u &= 
-\frac{\Mnn\hs (99 \!-\!8\hs\ct \!+\! 28\hs\ctt 
\!+\! 8\hs\cttt \!+\hsm \ctf)}
{8\hs (1\!+\!\ct)^2}\,.
\end{align}
\eeqs

Using the double-copy formulas in Eqs.\eqref{Amp-hL-nnnn}-\eqref{Amp-phi-nnnn} 
together with 
the gauge-transformed numerators
$(\NN_j^{\pp},\,\NNt_j^{\pp})$ 
in 
\eqrefe{eq:N0j'} 
and Eqs.\eqref{eq:Nj'-NLO}-\eqref{eq:Ntj'-NLO},
we construct the following 
four-point KK gravition amplitude and gravitational
KK Goldstone amplitude at the LO and NLO:    
\beqs
\begin{align}
\label{eq:DC-Amp-LOf}
\M_0^{}(\text{DC}) 
&=	\MT_0(\text{DC})= 
\frac{~3\ka^2\,}{~128~}\hs s\hs  
( 7\hsm +\ctt )^2\hsm \csc^2\!\theta  \,,
\\[2mm]
\label{eq:DC-Amp-hL-NLOf}
\da\M (\text{DC}) 
&= -\frac{~5\ka^2 M_n^2~}{768}
(1642 \hsm +\hsm  297\hs\ctt 
\hsm +\hsm 102\hs\ctf\hsm +\hsm 7\hs\cts)
\hsm \csc^4 \hsmx\theta  \,,
\\[2mm]
\label{eq:DC-Amp-phi-NLOf}
\da\MT (\text{DC}) 
&=- \frac{~\ka^2 M_n^2~}{768}   
(6386 \hsm +\hsm 3837 \hs\ctt 
\hsm +\hsm 30\hs\ctf 
\hsm -\hsm 13\hs\cts)\hsm\csc^4 \hsmx\theta\,,
\end{align}
\eeqs
where we have set the conversion constant 
$\,c_0^{}\!=\!-\ka^2/(24g^2)\hs$.\
The double-copy amplitudes of Eq.\eqref{eq:DC-Amp-LOf} 
provide the LO gravitational amplitudes 
\eqref{eq:ML0=M50}
and the NLO gravitational amplitudes 
\eqref{eq:dM-dmT-DC}
in the main text.
We can further compute the gravitational residual term of the GRET
from the difference between the two NLO amplitudes 
\eqref{eq:DC-Amp-hL-NLOf} and \eqref{eq:DC-Amp-phi-NLOf}:
\begin{equation}
\label{eq:Rterm-DC}
\Delta\M (\rm{DC}) 
\,=\, \da\M (\rm{DC}) - \da\MT (\rm{DC})
= -\ka^2 \Mnn \,( 7 + \ctt ) \hs,
\end{equation}
which provides 
\eqrefe{eq:Diff2-AmpDC-nnnn} 
in the main text.
We see that the above reconstructed residual term \eqref{eq:Rterm-DC}
by the extended double-copy approach 
does give the same size of $\mO(E^0\Mnn)$ 
and takes the same angular structure of $(1,\,\ctt)$ 
as the original residual term 
\eqref{eq:RTerm-NLO-GR} although their numerical coefficients 
still differ.\
As discussed in the main test, it is impressive to note that  
\eqrefe{eq:Rterm-DC} also demonstrates a very precise cancellation
between the angular structures 
$(1,\,\ctt,\,\ctf,\,\cts)\!\times\!\csc^4\!\theta\,$
of the NLO double-copied KK amplitudes 
\eqref{eq:DC-Amp-hL-NLOf}-\eqref{eq:DC-Amp-phi-NLOf}
down to the substantially simpler angular structure
$(1,\,\ctt)\hs$.\ 
This is the same kind of angular cancellations 
as what we found for the original NLO KK
graviton and Goldstone amplitudes 
\eqref{eq:Amp-E-2hLxx}-\eqref{eq:Amp-E-2phixx} and
their difference \eqref{eq:RTerm-NLO-GR}. 
This demonstrates that the above double-copied NLO KK amplitudes
have captured the essential features of the original KK
graviton (Goldstone) amplitudes at both the LO and NLO.
We have presented the further improved
NLO numerators \eqref{eq:dN'dNT'-XzzT}-\eqref{eq:sol-z-zT} 
in the main text,
which can realize the double-copied NLO KK amplitudes
in full agreement with the original NLO KK
graviton and Goldstone amplitudes 
\eqref{eq:Amp-E-2hLxx}-\eqref{eq:Amp-E-2phixx}.
A further study based on the first principle approach of
the KK string theory is recently presented in Ref.\,\cite{Li:2021yfk},
which can realize the exact double-copy construction
of the general $N$-point KK graviton scattering amplitudes 
at tree level.


\vspace*{1mm}

Finally, for the sake of comparison, we also give the results
of making the high energy expansion of $\Mnn/\sz\hs$
and explain that within this expansion there is no
generalized gauge transformation which could realize the
Jacobi-conserving numerators 
for KK gauge boson (Goldstone)
scattering amplitudes.\ 
For this, we express the elastic scattering amplitude
$\,\T[4A^{n}_{L}]\hsm\equiv\hsm
\T[A^{an}_L A^{bn}_L \hsm\ito\hsm A^{cn}_L A^{dn}_L]$ and
$\tT[4A^{n}_5] \hsm\equiv\hsm 
\tT[A^{an}_5 A^{bn}_5 \hsm\ito\hsm A^{cn}_5 A^{dn}_5]$
as follows:
\beqs
\label{Amp0-ALA5-nnnn}
\begin{align}
\label{Amp0-AL-nnnn}
\T[4A^{n}_{L}]
&\,=\, g^2\! \(\! \frac{\,\CC_s \NN_s\,}{\sz}
+ \frac{\,\CC_t \NN_t\,}{\tz}
+ \frac{\,\CC_u \NN_u\,}{\uz}  \!\)  \!,
\\[1mm]
\label{Amp0-A5-nnnn}
\tT[4A^{n}_5]
&\,=\, g^2\! \(\! \frac{\,\CC_s \NNt_s\,}{\sz}
+ \frac{\,\CC_t \NNt_t\,}{\tz}
+ \frac{\,\CC_u \NNt_u\,}{\uz}  \!\)  \!.
\end{align}
\eeqs
We compute their numerators at the LO and NLO,
$(\NN_j,\,\NNt_j) = (\NN_j^0,\,\NNt_j^0)
\hsm +\hsm (\da\NN_j,\,\da\NNt_j)
= \mO(E^2M_n^0)+\mO(E^0\Mnn)\hs$,
and present them in the following Tabel\,\ref{stab:1}.

\vspace*{1mm}

With these, we verify that the LO numerators of KK gauge boson (Goldstone) scattering amplitude satisfy the Jacobi identity:
\begin{equation}
\sum_j \NN^0_j = 0 \,, \qquad
\sum_j \NNt^0_j =0 \,,
\end{equation}
where $j \in (s,t,u)$\,.
But, we find that the Jacobi identity is
no longer obeyed by the NLO numerators:
\beqs
\begin{align}
& \sum_j \da \NN_j = \sum_j \da \NNt_j = \chi \neq 0\,,
\\
& ~\chi \,= -2\hs (7\!+\ctt)\hs\ct\csc^2\!\theta \,M_n^2 \,,
\end{align}
\eeqs
We further note that the KK amplitudes
\eqref{Amp0-AL-nnnn}-\eqref{Amp0-A5-nnnn}
are invariant under the generalized gauge transformations
for the kinematic numerators:
\begin{equation}
\NN_j \to \NN^{\pp}_j =\NN_j + \Delta \!\times\! s_{0j}^{} \,,
\hspace*{6mm}
\NNt_j \to \NNt^{\pp}_j =
\NNt_j + \widetilde{\Delta} \!\times\! s_{0j}^{} \,.
\end{equation}
But, because of $\,\sum_j\! s_{0j}^{}\!=0$\,
[cf.\,\eqrefe{eq:stu-stu0}], we deduce
$\,\sum_j\!\da\NN^{\pp}_j\!\hsm =\!\sum_j\da\NN_j^{}\!\neq\hsm 0\,$
and
$\,\sum_j\!\da\NNt^{\pp}_j\!\hsm =\!\sum_j\da\NNt_j^{}\!\neq\hsm 0\,$.
Hence, under the expansion of $\hs\Mnn /\sz\hs$,
it is impossible to obtain proper solutions of
$\{\Delta\,,\widetilde{\Delta}\}$ which are supposed to ensure the
gauge-transformed NLO numerators
$(\da\NN_j^{\pp},\,\da\NNt_j^{\pp}\hs)$
to obey the kinematic Jacobi identity.

{\linespread{1.85}
\begin{table*}[h]
\centering
\caption{%
\baselineskip 13pt
Kinematic numerators of the LO and NLO scattering amplitudes
\eqref{Amp0-AL-nnnn}-\eqref{Amp0-A5-nnnn}
for KK longitudinal gauge bosons and KK Goldstones
under the high energy expansion of $\Mnn/\sz\hs$, where
$(\NN_j,\,\NNt_j) = (\NN_j^0,\,\NNt_j^0) \!+\! (\da\NN_j,\,\da\NNt_j)
= \mO(E^2M_n^0)+\mO(E^0\Mnn)\hs$.}
\vspace*{1mm}
\begin{tabular}{c||c|c|c||c|c|c||c|c|c}
\hline\hline
~Numerators~ &$\quad \NN_s^{} \quad $  &$\NN_t^{}$  &$\NN_u^{}$  &$\quad\NNt_s^{}\quad$  &$\NNt_t^{}$  &$\NNt_u^{}$
&$\,\NN_s^{}\!-\!\NNt_s^{}\,$
&$\,\NN_t^{}\!-\!\NNt_t^{}\,$
&$\,\NN_u^{}\!-\!\NNt_u^{}\,$ \\
\hline
$\NN^0_j/\sz$
&\large$-\frac{\,11\ct\,}{2}$
&\large$\frac{\,-5+11 \ct + 4 \ctt\,}{4}$
&\large$ \frac{\,5+11 \ct - 4 \ctt\,}{4}$
&\large$-\frac{\,3\ct\,}{2} $
&\large$\frac{\,3(-3+\ct)\,}{4}$
&\large$\frac{\,3(3+\ct)\,}{4}$
&$-4 \ct$
&$-4 \ct$
&$-4 \ct$
\\ \hline
$\dNN_j^{}/\Mnn$
&$4\ct$
&\large$\frac{\,2 (2-3\ct-2\ctt-\cttt)\,}{1+\ct\,}$
&\large$-\frac{\,2 (2+3\ct-2\ctt+\cttt)\,}{1-\ct\,}$
&$\,4\ct\,$
&\large$-\frac{\,8\ct\,}{\,1+\ct\,}$
&\large$-\frac{\,8\ct\,}{\,1-\ct\,}$
&$0$
&$8 s_{\theta}^2$
&$-8 s_{\theta}^2$ \\
\hline\hline
\end{tabular}
\label{stab:1}
\end{table*}}


\begin{thebibliography}{99}
\vspace*{-4mm}

\bibitem{KK}
T.\ Kaluza,\ 
``On the Unification Problem in Physics'',
Sitzungsber.\ Preuss.\ Akad.\ Wiss.\ Berlin (Math. Phys.)
1921 (1921) 966 [Int.\ J.\ Mod.\ Phys.\ D\,27 (2014) 1870001,
[arXiv:1803.08616];
O.\ Klein,
``Quantum Theory and Five-Dimensional Theory of Relativity'',
Z.\ Phys.\ 37 (1926) 895
[Surveys High Energ.\ Phys.\ 5 (1986) 241].


\bibitem{Exd}
N.\ Arkani-Hamed, S.\ Dimopoulos, and G.\ R.\ Dvali,
Phys.\ Lett.\ B\,429 (1998) 263 [arXiv:hep-ph/9803315];
I.\ Antoniadis, N.\ Arkani-Hamed, S.\ Dimopoulos, and
G.\ R.\ Dvali, Phys.\ Lett.\ B\,436 (1998) 257 [arXiv:hep-ph/9804398];
L.\ Randall and R.\ Sundrum, Phys.\ Rev.\ Lett.\ 83 (1999) 3370
[arXiv:hep-ph/9905221]. 


\bibitem{string}
M.\ B.\ Green, J.\ H.\ Schwarz, and E.\ Witten,
``Superstring Theory'',
Cambridge University Press, 1987;
J.\ Polchinski, ``String Theory'',
Cambridge University Press, 1998.


\bibitem{GHiggs}
L.\ Dolan and M.\ Duff,
Phys.\ Rev.\ Lett.\ 52 (1984) 14;\\
Y.\ M.\ Cho and S.\ W.\ Zoh
Phys.\ Rev.\ D\,46 (1992) 2290.


\bibitem{5DYM2002}
R.\ S.\ Chivukula, D.\,A.\ Dicus, H.\ J.\ He,
Phys.\ Lett.\ B 525 (2002) 175 [hep-ph/0111016].


\bibitem{higgsM}
F.\ Englert and R.\ Brout, Phys.\ Rev.\ Lett.\ 13 (1964) 321;
P.\ W.\ Higgs, Phys.\ Rev.\ Lett.\ 13 (1964) 508;
Phys.\ Lett.\ 12 (1964) 132;
G.\ S.\ Guralnik, C. R. Hagen and T. Kibble,
Phys.\ Rev.\ Lett.\ 13 (1965) 585;
T. Kibble, Phys.\ Rev.\ 155 (1967) 1554.


\bibitem{vDVZ}
H.\ van Dam and M.\ J.\ G.\ Veltman,
Nucl.\ Phys.\ B\,22 (1970) 397;
%
V.\ I.\ Zakharov JETP Letters (Sov.~Phys.) 12 (1970) 312.


\bibitem{PF}
M.\ Fierz and W.\ Pauli,
Proc.\ Roy.\ Soc.\ Lond.\ A\,173 (1939) 211.

\bibitem{Hinterbichler:2012}
For a review, K.\ Hinterbichler,
Rev.\ Mod.\ Phys. 84 (2012) 671 [arXiv:1105.3735 [hep-th]].


\bibitem{5DYM2002-2}
R.\ S.\ Chivukula and H.\ J.\ He,
Phys.\ Lett.\ B 532 (2002) 121 [hep-ph/0201164].


\bibitem{KK-ET-He2004}
H.-J. He, Int.\ J.\ Mod.\ Phys.\ A\,20 (2005) 3362
[arXiv:hep-ph/0412113] (cf.\ its section\,3), and
presentation at DPF-2004: Annual Meeting of the Division of Particles
and Fields, American Physical Society, August 26-31, 2004,
Riverside, California, USA.


\bibitem{Chivukula:2019S}
R.\ S.\ Chivukula, D.\ Foren, K.\ A.\ Mohan, D.\ Sengupta, and E.\ H.\ Simmons,
Phys.\ Rev.\ D 101 (2020) 055013 
[arXiv:1906.11098 [hep-ph]].

\bibitem{Chivukula:2019L}
R.\ S.\ Chivukula, D.\ Foren, K.\ A.\ Mohan, D.\ Sengupta, and E.\ H. Simmons,
Phys.\ Rev.\ D 101 (2020) 075013
[arXiv:2002.12458 [hep-ph]].


\bibitem{Kurt}
J.\ Bonifacio and Kurt Hinterbichler,
JHEP 1912 (2019) 165 [arXiv:1910.04767 [hep-th]].


\bibitem{Elvang:2013}
For a review,
H.\ Elvang and Y.\ T.\ Huang, ``Scattering Amplitudes'',
[arXiv:{1308.1697} [hep-th]],
Cambridge University Press, 2015.


\bibitem{BCJ:2008}
Z.\ Bern, J.\ J.\ M.\ Carrasco, H.\ Johansson,
Phys.\ Rev.\ D 78 (2008) 085011
[arXiv:{0805.3993} [hep-th]];
Phys.\ Rev.\ Lett. 105 (2010) 061602
[arXiv:{1004.0476} [hep-th]].

\bibitem{BCJ:2019}
For a review, Z.\ Bern, J.\ J.\ M.\ Carrasco, M.\ Chiodaroli, H.\ Johansson, R.\ Roiban,
[arXiv:{1909.01358} [hep-th]].


\bibitem{KLT}
H.\ Kawai, D.\,C.\ Lewellen, and S.\,H.\,H.\ Tye,
Nucl.\ Phys.\ B\,269 (1986) 1-23.

\bibitem{Tye-2010}
S.\,H.\,H.\ Tye and Y.\ Zhang,
JHEP 1006 (2010) 071 [arXiv:\ 1003.1732 [hep-th]].

\bibitem{RS}
L.\ Randall and R.\ Sundrum,
Phys.\ Rev.\ Lett.\ 83 (1999) 3370 
[arXiv:hep-ph/9905221].


\bibitem{supp}
Y.~F.~Hang and H.~J.~He,
Supplemental Material.


\bibitem{Hang:2021fmp}
Y.~F.~Hang and H.~J.~He,
Phys.\ Rev.\ D {105} (2022) 084005 
arXiv:2106.04568 [hep-th].
 


\bibitem{ET94}
H.\ J.\ He, Y.\,P.\,Kuang and X.\,Li,
Phys.\ Rev.\ D 49 (1994) 4842;
Phys.\ Rev.\ Lett.\ 69 (1992) 2619.


\bibitem{GET-2}
Y.-F.~Hang and H.-J.~He, in preparation.


\bibitem{ET96}
H.\ J.\ He and W.\,B.\ Kilgore,
Phys.\ Rev.\ D 55 (1997) 1515 [hep-ph/9609326].

\bibitem{ET-Rev}
For a comprehensive review of the conventional ET in 4d,
H.\ J.\ He, Y.\,P.\ Kuang and C.\,P.\ Yuan,
arXiv:hep-ph/9704276 and DESY-97-056, in the proceedings of the workshop on ``Physics at the TeV Energy Scale'', vol.72, p.119
(Gordon and Breach, New York, 1996).


\bibitem{weinberg}
S.\ Weinberg, Physica 96A (1979) 327.


\bibitem{steve-foot}
We reposted an update version of the 
companion long paper\,\cite{Hang:2021fmp} to
arXiv:2106.04568v3 in which we presented our generalized 
power counting method for the compactified KK gauge theories 
and KK GR theories. But we learnt the sad news on the following day 
that Steven Weinberg passed away on July\,23.  
One of us (HJH) wishes to express his deep gratitude to Steve
for his inspirations over the years 
(including the discussion of his original work 
on the power counting rule\,\cite{weinberg}),
especially during his times at UT Austin where his office
was only a few doors away from that of Steve.


\bibitem{Li:2021yfk}
Y.~Li, Y.-F.~Hang, H.-J.~He, and S.~He,
JHEP 02 (2022) 120
[arXiv:2111.12042 [hep-th]].


\bibitem{5dSM}
R.\ S.\ Chivukula, D.\ A.\ Dicus, H.\ J.\ He, and S.\ Nandi,
Phys.\ Lett.\ B 562 (2003) 109 [hep-ph/0302263].


\bibitem{CHY}
F.~Cachazo, S.~He, E.~Y.~Yuan,
Phys.\ Rev.\ D {90} (2014) 065001 
[arXiv:1306.6575 [hep-th]];
Phys.\ Rev.\ Lett.\ {113} (2014) 171601 
[arXiv:1307.2199 [hep-th]];\
%
JHEP {1407} (2014) 033
[arXiv:1309.0885 [hep-th]].


\bibitem{Hang:2021oso}
Y.-F.~Hang, H.-J.~He, and C.~Shen,
JHEP 01 (2022) 153 [arXiv:2110.05399 [hep-th]].

\bibitem{TMG} 	
S.~Deser, R.~Jackiw, and S.~Templeton,
Phys.\ Rev.\ Lett.\ 48 (1982) 975;
Annals Phys. 140 (1982) 372-411.


\end{thebibliography}
\end{document}